\documentclass[pdflatex,sn-mathphys-num]{sn-jnl}

\usepackage{graphicx}%
\usepackage{multirow}%
\usepackage{amsmath,amssymb,amsfonts}%
\usepackage{amsthm}%
\usepackage{mathrsfs}%
\usepackage[title]{appendix}%
\usepackage[table]{xcolor}
\usepackage{textcomp}%
\usepackage{manyfoot}%
\usepackage{booktabs}%
\usepackage{algorithm}%
\usepackage{algorithmicx}%
\usepackage{algpseudocode}%
\usepackage{listings}%

\usepackage{multirow}
\usepackage{booktabs}
\usepackage{subcaption}
\usepackage{hyperref}
\usepackage{mathrsfs}
\usepackage{stmaryrd}
\usepackage{tikz}
\usetikzlibrary{shapes.geometric, arrows, calc}
\usepackage{array}

\theoremstyle{definition}%
\newtheorem{definition}{Definition}%

\newcommand{\rss}[1]{\ensuremath{\llbracket{#1}\rrbracket}}
\newcommand{\rssa}[1]{\ensuremath{\rss{#1}^\mathsf{A}}}
\newcommand{\rssb}[1]{\ensuremath{\rss{#1}^\mathsf{B}}}

\newcommand{\Enc}{\ensuremath{\mathsf{Enc}}}
\newcommand{\Dec}{\ensuremath{\mathsf{Dec}}}
\newcommand{\ct}{\ensuremath{\mathsf{CT}}}
\newcommand{\ctsymm}{\ensuremath{\mathsf{ct}}}
\newcommand{\ad}{\ensuremath{\mathit{ad}}}
\newcommand{\pk}{\ensuremath{\mathsf{PK}}}
\newcommand{\sk}{\ensuremath{\mathsf{SK}}}
\newcommand{\st}{\ensuremath{\mathsf{st}}}

\DeclareMathOperator{\FHEEnc}{Enc_{CKKS}}
\DeclareMathOperator{\FHEDec}{Dec_{CKKS}}

\DeclareMathOperator{\FHERot}{Rot}
\DeclareMathOperator{\FHEAdd}{Add}
\DeclareMathOperator{\FHEMul}{Mul}

\newcommand{\RR}[0]{\ensuremath{\mathbb{R}}}
\newcommand{\ZZ}[0]{\ensuremath{\mathbb{Z}}}

\begin{document}

\title[MOZAIK: A Privacy-Preserving Analytics Platform for IoT Data Using MPC and FHE]{MOZAIK: A Privacy-Preserving Analytics Platform for IoT Data Using MPC and FHE}


\author*[1]{\fnm{Michiel} \spfx{Van} \sur{Kenhove}}\email{michiel.vankenhove@ugent.be}
\equalcont{These authors contributed equally to this work.}

\author[2]{\fnm{Erik} \sur{Pohle}}\email{erik.pohle@esat.kuleuven.be}
\equalcont{These authors contributed equally to this work.}

\author[2]{\fnm{Leonard} \sur{Schild}}\email{leonard.schild@esat.kuleuven.be}
\equalcont{These authors contributed equally to this work.}

\author[2]{\fnm{Martin} \sur{Zbudila}}\email{martin.zbudila@esat.kuleuven.be}
\equalcont{These authors contributed equally to this work.}

\author[1]{\fnm{Merlijn} \sur{Sebrechts}}\email{merlijn.sebrechts@ugent.be}

\author[1]{\fnm{Filip} \spfx{De} \sur{Turck}}\email{filip.deturck@ugent.be}

\author[1]{\fnm{Bruno} \sur{Volckaert}}\email{bruno.volckaert@ugent.be}

\author[2]{\fnm{Aysajan} \sur{Abidin}}\email{aysajan.abidin@esat.kuleuven.be}

\affil*[1]{\orgdiv{IDLab}, \orgname{Ghent University - imec, Department of Information Technology}, \orgaddress{\street{Technologiepark-Zwijnaarde 126}, \city{Ghent}, \postcode{9052}, \country{Belgium}}}

\affil[2]{\orgdiv{COSIC}, \orgname{KU Leuven, Department of Electrical Engineering}, \orgaddress{\street{Kasteelpark Arenberg 10}, \city{Leuven-Heverlee}, \postcode{3001}, \country{Belgium}}}



\abstract{The rapid increase of Internet of Things (IoT) systems across several domains has led to the generation of vast volumes of sensitive data, presenting significant challenges in terms of storage and data analytics. Cloud-assisted IoT solutions offer storage, scalability, and computational resources, but introduce new security and privacy risks that conventional trust-based approaches fail to adequately mitigate. To address these challenges, this paper presents MOZAIK, a novel end-to-end privacy-preserving confidential data storage and distributed processing architecture tailored for IoT-to-cloud scenarios. MOZAIK ensures that data remains encrypted throughout its lifecycle, including during transmission, storage, and processing. This is achieved by employing a cryptographic privacy-enhancing technology known as computing on encrypted data (COED). Two distinct COED techniques are explored, specifically secure multi-party computation (MPC) and fully homomorphic encryption (FHE). The paper includes a comprehensive analysis of the MOZAIK architecture, including a proof-of-concept implementation and performance evaluations. The evaluation results demonstrate the feasibility of the MOZAIK system and indicate the cost of an end-to-end privacy-preserving system compared to regular plaintext alternatives. All components of the MOZAIK platform are released as open-source software alongside this publication, with the aim of advancing secure and privacy-preserving data processing practices.}

\keywords{Internet of Things (IoT), Secure Data Collection, Privacy-Preserving Machine Learning Inference, Secure Multi-Party Computation (MPC), Fully Homomorphic Encryption (FHE), Data Intermediaries}



\maketitle

\section{Introduction}\label{sec:introduction}

The pervasive deployment of versatile Internet of Things (IoT) systems in daily life, such as in healthcare~\cite{selvaraj2020challenges}, smart cities~\cite{latre_city_2016,santos_city_2018}, and smart homes~\cite{ansari_smart_2024}, continuously generate vast amounts of high-definition images, videos, and sensor data. Massive IoT data demands substantial storage and high-performance computation power, often exceeding the capabilities of typical users or IoT devices~\cite{bautista_mpc-as-service_2023}. To address this, cloud-assisted IoT leverages the cloud's computational and storage capabilities, providing additional conveniences as a data intermediary. However, this added convenience also introduces new security risks, including data loss and unauthorized access, which present significant challenges in developing secure cloud-assisted IoT systems. Conventional security mechanisms~\cite{mosenia_comprehensive_2017} are insufficient to address these risks, and trust-based~\cite{najib_trust_2022,hosseini_shirvani_survey_2023} approaches fall short in providing adequate security, exposing users to privacy risks and companies to liability and reputational risks. Additionally, as data intermediaries, companies must fulfill their obligations under the Data Governance Act (DGA) while also adhering to the General Data Protection Regulation (GDPR) when handling personal data from users inside the European Union. Moreover, ensuring privacy through data anonymization and obfuscation in IoT environments proves impractical, particularly for dynamic operations such as insertion and deletion. This challenge becomes even more pronounced when outsourcing storage or computation to the cloud, where these privacy measures fail to provide provable security~\cite{wang_secure_2018}. Consequently, the absence of secure and reliable data-sharing and analytics platforms, coupled with the lack of privacy-compliant analytics for personal and proprietary data, hinders the growth of a data-driven market and economy. In turn, this significantly restricts data sharing and technological innovation.

To tackle these challenges, the MOZAIK research project~\cite{noauthor_mozaik_project_nodate} proposes an end-to-end confidential data storage and distributed processing solution for IoT-to-cloud scenarios. This paper presents an in-depth exploration and evaluation of MOZAIK's novel architecture, which ensures that data from IoT devices is encrypted before leaving the user's control, remains encrypted while being securely stored in an IoT-data-focused cloud-based data store, and is processed using privacy-preserving machine learning inference without revealing the plaintext to any single entity in the system. To achieve this, MOZAIK leverages privacy-enhancing technologies (PETs), enabling data owners to securely share sensitive data with data processors without compromising privacy. While real-world systems often rely on statistical PET techniques such as differential privacy, synthetic data generation, and federated learning, MOZAIK employs cryptographic PET techniques. More specifically, computing on encrypted data (COED), that offers robust cryptographic security guarantees to data owners. MOZAIK integrates two key COED techniques, multi-party computation (MPC) and fully homomorphic encryption (FHE), ensuring end-to-end secure privacy-preserving data processing. To the best of our knowledge, MOZAIK is the first fully end-to-end confidential IoT data storage and privacy-preserving processing platform.


This paper substantially expands upon two of our previously published conference papers~\cite{abidin_mozaik_2023,marquet_secure_2023}, providing an in-depth and detailed analysis of the MOZAIK system, presenting significant advancements and novel evaluation results and insights. It includes notable progress and improvements made to the system's architecture, and a proof-of-concept implementation of all its components.
We further provide both end-to-end and micro-benchmark evaluations to assess the performance and overhead of the proposed end-to-end privacy-preserving confidential data storage and processing solution. Specifically, our contributions in this paper are as follows:

\begin{itemize}
    \item Design a novel system architecture to enable scalable and end-to-end secure privacy-preserving computation and analytics on sensitive IoT data.
    \item Demonstrate the system's feasibility by applying a specific real-life use case as an implementation on top of the system's architecture.
    \item Evaluate the performance overhead of the designed system in comparison to non-privacy-preserving processing techniques by conducting end-to-end benchmarks and micro-benchmarks of specific system components.
    \item Provide the complete MOZAIK platform, its implementation, and all its components as open-source software in the following GitHub organization~\cite{van_kenhove_github_nodate}.
\end{itemize}

The remainder of this paper is structured as follows. We begin by positioning our work within the existing literature in Section~\ref{sec:related-work}. Then, we introduce key concepts and the required background, including mathematical notation, in Section~\ref{sec:background}. Section~\ref{sec:design} describes the design and implementation of the system's architecture, detailing the functionality of each component and how they communicate with each other. In Section~\ref{sec:use-case}, we introduce a specific use case that was implemented as a practical proof of concept to validate MOZAIK's architecture design. The implementation details specific to this use case are described in Section~\ref{sec:implementation}. In Section~\ref{sec:evaluation}, we assess the performance and overhead of the system using both end-to-end benchmarks and micro-benchmarks. Finally, we conclude the paper in Section~\ref{sec:conclusion}.

\section{Related Work}\label{sec:related-work}

A broad spectrum of platforms has been proposed for privacy-preserving data sharing and processing, particularly in contexts of healthcare, edge devices, and decentralized marketplaces. These works often address only isolated phases of the data lifecycle or rely on centralized trust models, limiting their applicability in fully privacy-preserving systems.

The ``MHMD: My Health, My Data'' project~\cite{MHMD/DBLP:conf/edbt/Morley-Fletcher17} and the work by Shafagh et al.~\cite{DBLP:conf/ccs/ShafaghBHD17} both present a distributed ledger-based platform focusing on access control features and data management for medical and IoT data, respectively. However, neither system offers mechanisms for protecting the privacy of the underlying data during processing. In contrast, our work explicitly prioritizes data confidentiality, ensuring end-to-end secure data processing while the user maintains control over their data throughout its lifecycle.

A more privacy-focused approach is taken by Agora~\cite{Agora/DBLP:journals/tdsc/KoutsosPCTH22}, which leverages functional encryption to achieve input privacy during smart contract execution on blockchain. However, Agora depends on a trusted third party for key generation, and like many such systems, it does not address the problem of secure data ingestion or long-term storage privacy. Niu et al.~\cite{DBLP:journals/tkde/NiuZWGC19} similarly employ partially homomorphic encryption to enable privacy-preserving data markets, but their reliance on a central authority holding the decryption keys weakens output privacy and introduces a single point of trust. In our work we propose an end-to-end secure data processing platform based on cryptographic techniques, removing the need for a central entity with access to any unencrypted plaintext data.

Several projects use secure multi-party computation to enable collaborative, privacy-preserving processing. The KRAKEN platform~\cite{Kraken/DBLP:books/sp/22/GabrielliKPBBRV22,DBLP:conf/primelife/KochKPR20} adopts MPC to protect data during processing. Likewise, Veeningen et al.~\cite{DBLP:conf/mie/VeeningenCHSBSG18} present an MPC-based system for secure medical analysis, supplemented by GDPR compliance considerations. However, they rely on intermediate entities, such as hospitals or general practitioners, where data is stored or transmitted in the clear prior to computation, thus leaving the early stages of the data pipeline unprotected. Januszewicz et al.~\cite{DBLP:journals/iacr/JanuszewiczGKZTLJ24} explore secure aggregation in streaming scenarios using functional encryption-like techniques, but their work focuses on a narrow computation model and does not incorporate secure data collection.

A more comprehensive privacy model is discussed by Golob et al.~\cite{DBLP:conf/de/GolobPDDLCN23}, who combine MPC with differential privacy to protect both input and output privacy during processing in a decentralized marketplace. Nevertheless, their platform does not specify secure data collection mechanisms and remains focused on processing in isolation.

With regard to secure computation over encrypted data, a tremendous number of works have recently emerged, offering solutions across various deployment settings~\cite{zbudila_sok_2025,ng_sok_2023}. Many MPC-based solutions target the outsourcing setting, where inputs are distributed among computing parties performing machine learning (ML) inference. While often more efficient, such works typically lack integration with data providers. Another line of research considers the client-server scenario, in which a client holding the query and a server owning the ML model engage in secure computation directly, typically via FHE or hybrid FHE-MPC schemes~\cite{zbudila_sok_2025,ng_sok_2023}. However, these approaches include the client device as part of the computation cluster, thereby imposing undesirable overhead on the client side. Our work integrates these approaches by leveraging the more efficient outsourced secure processing and tackling the data lifecycle end-to-end, from collection and storage to processing, while minimizing the overhead on the client-edge devices.

Several works focus specifically on distributed computation involving IoT devices and cloud servers. CryptDNN~\cite{li_cryptdnn_2025} presents a cloud-edge-client architecture based on a hybrid model combining fully homomorphic encryption with searchable symmetric encryption. The authors focus particularly on the efficiency and accuracy of the underlying computation primitives but provide limited analysis of the complete data lifecycle. In contrast, our work emphasizes a scalable, end-to-end secure, privacy-preserving storage and analytics platform for sensitive IoT data.

Similarly, the work of Yang et al.~\cite{yang_privacy-preserving_2025} is set in the cloud-edge setting under a hybrid FHE-MPC model combined with differential privacy. Their focus is ML training: an edge server performs shallow training on encrypted data using FHE, after which cloud servers combine models via weighted averaging to form an aggregated model. This aggregated model is then used to perform learning over secret shares to obtain the final target model. While their approach supports privacy-preserving model learning, it does not address secure data collection, storage, or privacy-preserving analysis via model inference. Our approach, by design, does not train models, but focuses on end-to-end secure collection, storage, and analysis of sensitive data using a pre-trained model supplied by a model provider.

PrivESD~\cite{wang_privesd_2025} proposes a privacy-preserving logistic regression framework in a cloud-edge collaborative architecture, targeting encrypted streaming data. It leverages multi-key fully homomorphic encryption to enable computation over data encrypted under different keys, mitigating collusion risks. By distributing computation between edge nodes and the cloud, PrivESD reduces computational complexity and communication overhead while maintaining confidentiality. However, it primarily targets privacy-preserving model training on streaming data. Our work, instead, supports continuous analysis of encrypted sensitive streams by enabling privacy-preserving model inference on individual data points or small micro-batches as they arrive.

In contrast to the above works, our approach addresses privacy and security across the entire data lifecycle, from the moment data is generated and collected, through encrypted transmission and privacy-preserving processing, to secure and verifiable storage. Unlike systems that isolate data protection to a single phase or depend on trusted third parties, our architecture ensures that no stage of the pipeline is exposed to unprotected handling or centralized control. Furthermore, our design effectively minimizes the computational and communication overhead on client-edge devices through the use of lightweight cryptographic schemes. This end-to-end, cryptographically grounded design provides a scalable and secure framework suitable for real-world deployments in untrusted environments, filling a critical gap left by existing solutions.

\section{Background}\label{sec:background}

This section provides essential background information and key concepts to help the reader better comprehend the continuation of the paper. To begin, we briefly familiarize the reader with the various notations used throughout this paper. Afterwards, we provide background information on data processing platforms, secure multi-party computation, and fully homomorphic encryption.

\subsection{Notation}
For this paper we use the notation of $P_i\in\mathcal{P}$ for party $i$. Let $\ell, n > 1$, we denote the ring of integers modulo $\ell$ with $\mathbb{Z}_{2^\ell}$ and let $GF(2^n)$ be a field extension of $\mathbb{Z}_2$. We use $x \| y$ to denote concatenation of bits/strings. We denote the uniformly random sampling from a finite set $A$ between parties $i$ and $j$ as $\xleftarrow{\$_{i,j}} A$. Such sampling is implemented using pseudorandom functions (PRFs), as described by Furukawa et al.~\cite{Furukawa2017}. 
We refer to an MPC protocol that performs a functionality $f$ as $\Pi_f$.

We briefly recall the syntax and semantics of symmetric and asymmetric \emph{authenticated} encryption. Informally, a nonce-based symmetric authenticated encryption scheme consists of two algorithms:

\begin{itemize}
    \item $\ctsymm_m^k \gets \Enc_k(m, N, \mathit{ad})$: Encrypts a message $m$ under the secret key $k$ with associated data $\mathit{ad}$ and nonce $N$, and produces a ciphertext $\ctsymm_m^k$.
    \item $m / \bot  \gets \Dec_k(\ctsymm_m^k, N, \mathit{ad})$: Decrypts the ciphertext $\ctsymm_m^k$ with the secret key $k$ and associated data $\mathit{ad}$. This operation only returns the original message $m$ if $\ctsymm_m^k$ was encrypted under the same $(k, N, \mathit{ad})$. Otherwise, it returns failure ($\bot$).
\end{itemize}

Similarly, an asymmetric authenticated encryption scheme for a private-public key pair $(\sk, \pk)$ consists of:

\begin{itemize}
    \item $\ctsymm_m^\pk \gets \Enc_{\pk}(m, \mathit{ad})$: Encrypts a message $m$ under the public key $\pk$ with associated data $\mathit{ad}$ and produces a ciphertext $\ctsymm_m^\pk$.
    \item $m / \bot  \gets \Dec_{\sk}(\ctsymm_m^\pk, \mathit{ad})$: Decrypts the ciphertext $\ctsymm_m^\pk$ with the private key $\sk$ and associated data $\mathit{ad}$. This operation only returns the original message $m$ if $\ctsymm_m^\pk$ was encrypted under the same $\pk$ and $\mathit{ad}$. Otherwise, it returns failure ($\bot$).
\end{itemize}

In the remaining part of this paper, we will always denote a symmetric secret key as $k$ and public/private keys as $\pk$/$\sk$. We use the notation $\ctsymm_m^K$ to denote the ciphertext of message $m$ under the key $K$, where $K$ is either a public key or a symmetric key.

\subsection{Data Processing Platforms}

Large-scale data is everywhere, with significant data growth in both commercial and scientific databases due to advancements in data generation and collection technologies. Data is gathered with the expectation that it will hold value, either for its intended purpose or for unforeseen future uses. High-quality data has become even more valuable with the recent advances in artificial intelligence (AI) and large language models (LLMs)~\cite{zha_data-centric_2025}. It is clear that simply storing data is insufficient to unlock its full potential value. To extract valuable insights, data must be actively processed. This need has driven the development of data processing platforms~\cite{singh_survey_2014}. These platforms are designed to efficiently manage, process, store, and analyze data from various sources. They enable us to make data-driven decisions, improve overall operational efficiency, and reduce costs. A data processing platform typically integrates multiple components and leverages various technologies to handle the entire data lifecycle.

To ensure the effectiveness of data processing platforms, they need to be reliable, scalable, and maintainable. The system should maintain its essential performance levels despite of adversity, such as hardware or software faults, or even human mistakes. As the load on the system increases, whether due to larger data volumes, higher traffic, or greater complexity, there should be reasonable methods to scale the system to maintain the desired performance without significantly increasing costs.

Data processing can be accomplished through batch processing, stream processing, or a combination of both. Batched processing involves the collection and processing of data in large predefined groups. This method is particularly effective for handling tasks that do not require immediate processing results. Batch processing systems operate by accumulating data over a period of time and start processing it all during off-peak times, thereby optimizing the use of computing resources. Stream processing, on the other hand, is used when near real-time results are required. A stream processor operates on each data point shortly after they are ingested in the system, whereas a batch job processes a fixed set of input data. Stream processing systems generally have lower latency to process a single data point compared to batch systems, but typically offer lower throughput than batch systems. A hybrid approach, known as micro-batch processing, tries to find a balance between latency and throughput. Micro-batch processing accumulates a smaller amount of data inputs and groups them in a batch before processing them. This improves the throughput of the data processing platform at the cost of a higher latency.

\subsubsection{Obelisk}\label{sec:background:obelisk}
Obelisk is a cloud-based time series database (TSDB) and storage platform designed and developed by researchers at the IDLab research group of Ghent University-imec. It is specifically designed for scalable IoT data ingestion and querying. Obelisk allows secured data ingestion, storage, streaming and retrieval through HTTPS REST (REpresentational State Transfer) APIs. With regards to data transport, Obelisk ensures server authentication, data confidentiality, and integrity towards its clients (IoT devices, computation servers, end users, etc.) against third-party adversaries through the use of the Transport Layer Security (TLS) protocol, a proven and widely adopted cryptographic protocol designed to secure communication over a computer network. Obelisk has been designed and implemented using an asynchronous, event-based, microservice software architecture, intended to be deployed on a container orchestration platform such as Kubernetes~\cite{noauthor_kubernetes_nodate}. The microservice-based architecture enables Obelisk to scale individual components under load horizontally, improving reliability and performance~\cite{bracke_design_2021}.

In order to contextualize our proposed software architecture in the remainder of the paper, we include a detailed overview of Obelisk HFS (High-Frequency Streaming). Although Obelisk HFS itself is not a novel contribution of this work, it serves as a foundational component within our system design.

\begin{figure}
    \centering
    \includegraphics[width=1.0\linewidth]{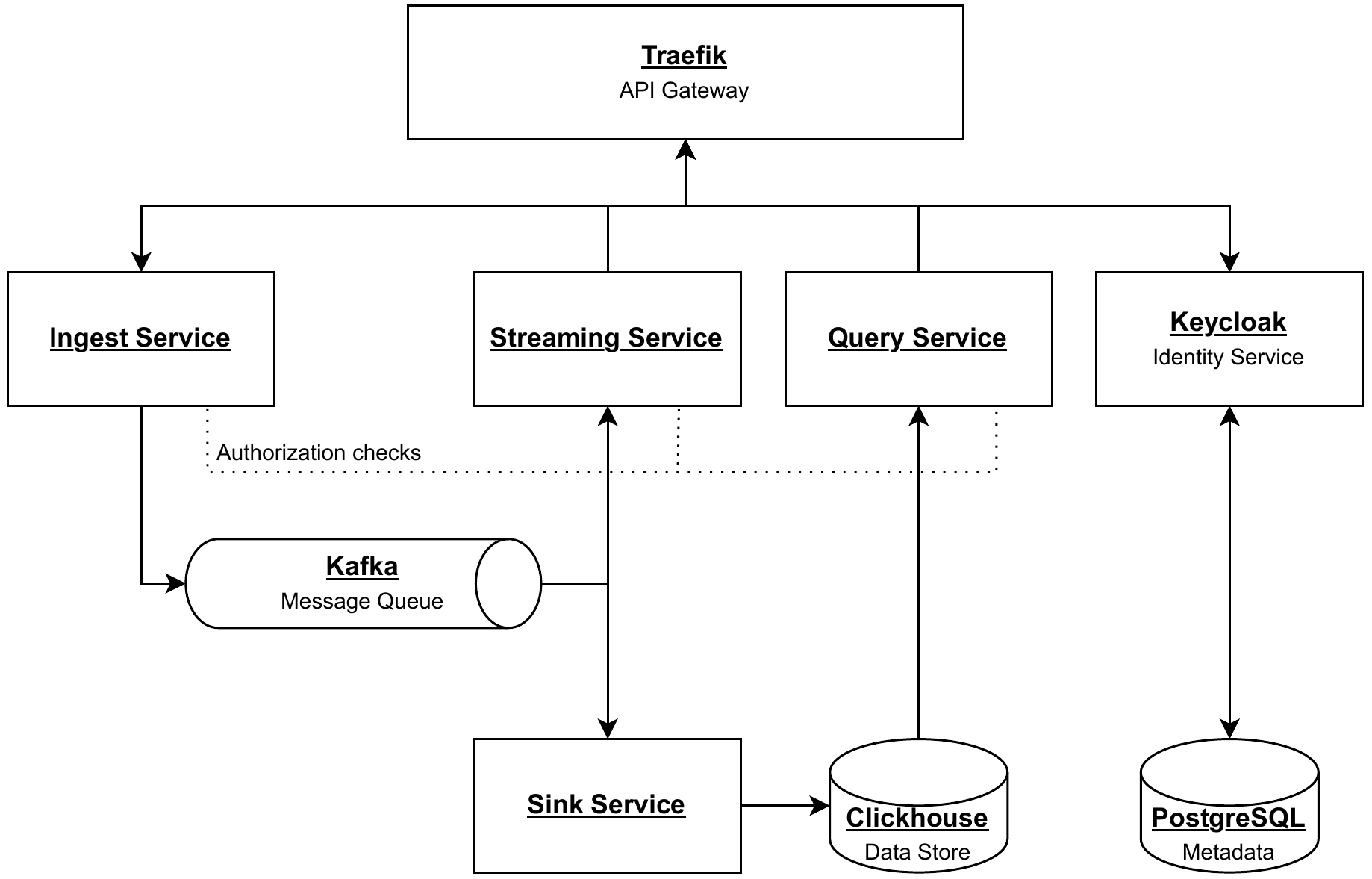}
    \caption{Architectural overview of Obelisk HFS. The system consists of five distinct microservices: the ingest service, the sink service, the streaming service, the query service, and Keycloak as the identity service. Ingested data is placed on a Kafka message queue topic to be consumed by the sink and streaming service. Data is eventually stored in a Clickhouse data store. The Traefik API gateway ensures requests are directed to the correct service.}\label{fig:obelisk_hfs_architecture}
\end{figure}

Obelisk HFS is a version of Obelisk specifically designed for high throughput. It has role-based access control (RBAC) built in, where specific permissions or dataset access needs to be explicitly granted to each entity interacting with the platform. Obelisk HFS consists of several microservice-based components, illustrated in the schematic overview of Figure~\ref{fig:obelisk_hfs_architecture}. The functionality of each component will be briefly summarized below.

\paragraph{API Gateway}
All requests made to Obelisk HFS arrive at its API gateway, which delegates each request to the correct service based on the HTTP URI of the request. For example, API requests with a URI of \texttt{/ingest} will be directed to Obelisk's ingest service, whereas requests with a URI or \texttt{/query} will be directed to the query service. The API gateway uses Traefik~\cite{sharma_traefik_2021} that serves as the ingress controller~\cite{noauthor_kubernetes_ingress_controller_nodate,noauthor_kubernetes_ingress_nodate} in a Kubernetes cluster.

\paragraph{Ingest Service and Message Queue}
The ingest service is responsible for data ingestion. To handle high-frequency data ingestion, only minimal processing is performed on each data point. Instead, the raw data is immediately published to an Apache Kafka~\cite{noauthor_apache_kafka_nodate} message queue topic, and a 204 HTTP response is returned to the data producer to acknowledge that it successfully received the data point. However, this acknowledgment does not guarantee that the data point has been stored in Obelisk's data store, as storage is handled asynchronously by the sink service and may require a significant amount of additional time. Ingesting data is thus an eventual-consistent operation. The rationale behind this design is to minimize the time required to acknowledge receipt of a data point, ensuring that the data producer is not unnecessarily blocked. Additionally, using a message queue for incoming data enables support for streaming use cases.

\paragraph{Sink Service and Data Store}
The sink service consumes data from the Kafka topic containing ingested data and ensures that these data points are reliably stored in the ClickHouse~\cite{clickhouse_clickhouse_nodate} data store. ClickHouse is a high-performance column-oriented database management system designed for online analytical processing. It excels at handling large volumes of data with fast query execution, making it well suited for real-time analytics and time-series data.

\paragraph{Query Service}
Once ingested data points are stored in ClickHouse, they are available to query through the query service, which allows the filtering and retrieval of data from datasets. As mentioned above, since data points are initially placed on a message queue before being stored by the sink service, there is a delay between ingestion and when the data becomes queryable.

\paragraph{Streaming Service}
Similar to the sink service, the streaming service functions as an additional consumer of the Kafka topic to which ingested data points are published. Rather than persisting the data in ClickHouse, the streaming service selectively forwards each data point that satisfies predefined criteria to subscribed clients using server-sent events~\cite{noauthor_html_server_sent_events_nodate}. Note that the sink service continues to consume all data points for storage, regardless of whether they are processed by the streaming service, ensuring eventual consistency. When a client creates a stream, they provide specific criteria, such as receiving only data points ingested in a certain dataset or those exceeding a threshold value. This selective forwarding requires real-time processing of each data point by the streaming service, thereby increasing system load relative to traditional poll-based querying. Nevertheless, it facilitates low-latency, real-time data delivery, which is essential for time-sensitive applications.

\paragraph{Identity Service}
The identity service in Obelisk is responsible for securely authenticating clients. To achieve this, Obelisk uses Keycloak~\cite{team_keycloak_nodate}, an open-source identity and access management solution that provides a secure and low-effort authentication system. In addition to authentication, Keycloak supports RBAC and more fine-grained authorization features. However, in Obelisk HFS, Keycloak is used solely for authentication, while authorization is managed by each service individually through configuration files. This approach will be replaced in an upcoming version of Obelisk, where fine-grained authorization will be handled using OpenFGA~\cite{noauthor_openfga_nodate}.

\subsection{Secure Multi-Party Computation}\label{sec:background:mpc}
Secure multi-party computation (MPC) is a cryptographic technique allowing multiple parties to jointly compute a public function on their corresponding data, without revealing their private inputs to any other parties. MPC protocols are designed to be secure in the presence of a certain number of corrupted parties. The set of corrupted parties denoted by $\mathcal{A}$ is assumed to be controlled by an adversary and may collude. Let $\mathcal{P}$ be the set of the computing parties. If the number of corrupted parties an MPC protocol tolerates is $|\mathcal{A}|<\frac{|\mathcal{P}|}{2}$, we say the MPC protocol is secure in the honest majority setting. Otherwise, if the protocol is able to tolerate $\frac{|\mathcal{P}|}{2}\leq |\mathcal{A}| < |\mathcal{P}|$, we say the MPC protocol is secure in the dishonest majority setting. We further differentiate protocols' security based on adversarial powers. A semi-honest, or passive, adversary is not allowed to deviate from the protocol. Such adversary strictly follows the protocol description but is curious to infer additional information from the data from their view. A malicious, or active, adversary can arbitrarily deviate from the protocol, thus manipulate any values or abort the execution at any time.

Many MPC frameworks work in the offline-online paradigm, where the computation can be split into two phases. The parties first engage in the input-independent preprocessing phase (sometimes called the offline phase), generating correlated randomness needed for the online phase. The parties can then run the online phase efficiently executing only input-dependent operations.

When realizing privacy-preserving machine learning (PPML) inference using MPC, we consider two types of approaches.
In a 2-party computation (2PC), the client represents one MPC party with its private data sample to be queried, and the other MPC party represents the ML server holding the trained ML model. Using 2PC, the parties can compute the inference without revealing their inputs to each other. This is often impractical due to the protocol overhead imposed on the client device. Therefore, many recent approaches to PPML inference involve outsourcing the computation to multiple independent third parties. In this approach, both the client and the model owner split their inputs among the predetermined computing parties, who then compute the ML inference using MPC. This ensures secure computation without revealing any information about the private inputs.

There are various cryptographic primitives that can enable MPC. For details regarding generic MPC primitives, we refer the reader to Evans et al.~\cite{SMPC/DBLP:journals/ftsec/EvansKR18}. In MOZAIK, the MPC protocols are based on a secret sharing technique. In secret sharing, each party with a private input splits its secret into multiple parts and distributes these among the computing parties.

\subsubsection{Secret Sharing}\label{sec:background:mpc:rss}
In this work we will mainly employ replicated secret sharing (see Definition~\ref{def:background:RSS}) either over the ring $\mathbb{Z}_{2^\ell}$, for $\ell \in \mathbb{N}$, or over the binary extension field $GF(2^n)$, for $n \in \mathbb{N}$.

\paragraph{Replicated Secret Sharing}
We use replicated secret sharing (RSS) as a 2-out-of-3 secret-sharing technique described by Furukawa et al.~\cite{Furukawa2017}. The secret is split among three parties, such that any two parties can reconstruct the secret. Note that protocols employing RSS are thus secure in the honest majority setting tolerating up to one corrupted party. We use two variants of replicated secret sharing in this paper: arithmetic sharing over the ring $\mathbb{Z}_{2^\ell}$ denoted as $\rssa{\cdot}$ and boolean sharing over the field $GF(2^n)$ denoted as $\rssb{\cdot}$.
\medskip

\begin{definition}[Replicated Secret Sharing]\label{def:background:RSS}
    Let $x\in\mathbb{Z}_{2^\ell}$. We denote the arithmetic replicated secret sharing of $x$ as $\rssa{x}$, where $x \equiv x_1 + x_2 + x_3 \mod 2^\ell$. 
    Let $x\in GF(2^n)$. We denote the boolean replicated secret sharing of $x$ as $\rssb{x}$, such that $x = x_1 \oplus x_2 \oplus x_3 $. 
    We define the share of party $P_i$ to be $\rss{x}_i = \left(x_i, x_{\left(1+i\right)\mod3}\right)$. We will use the notation $\rss{x}$ (without subscript) to denote operations on a RSS-shared $x$ performed by every party $i$.  
\end{definition}

\paragraph{Multiplication}
To obtain RSS shares of the product of two RSS-shared values $x$ and $y$, we make use of Beaver's trick~\cite{beaver_efficient_1992}, which works as follows.
The parties preprocess shares of a multiplication triple $\rss{a}, \rss{b}, \rss{c}$, such that $c=a\cdot b$.
In the online phase, the parties hold $\rss{x}, \rss{y}$ and wish to obtain $\rss{z}$, where $z = x \cdot y$. For this, the parties compute and reveal to all parties $d = x - a$ and $e = y - b$, and then set
\[ \rss{z} =  d\rss{b} + e\rss{a} + \rss{c} + de \,. \]

\paragraph{Conversion between $\rssa{x}$ and $\rssb{x}$}
In the following, we present two protocols to convert a value $x$ shared in arithmetic RSS, i.e., the parties hold $\rssa{x}$, into a boolean RSS, i.e., the parties obtain $\rssb{x}$. Afterwards, the reverse operation, converting from boolean to arithmetic, is presented.
We denote with $\mathsf{A2B}$ the arithmetic to boolean, and with $\mathsf{B2A}$ the boolean to arithmetic share conversion.
To efficiently convert between the two different types of secret sharing schemes, we employ a conversion protocol based on the ideas by Mohassel and Rindal~\cite{mohassel_aby3_2018}.
The conversion protocols use a Ripple Carry Adder (RCA) circuit, defined in Definition~\ref{def:design:RCA} below. \medskip

\begin{definition}[Ripple Carry Adder]\label{def:design:RCA}
    A Ripple Carry Adder is a circuit used for adding binary numbers. Let $x,y \in \mathsf{GF}(2^\ell)$, then the sum $x + y$ over $\mathbb{Z}_{2^\ell}$ can be computed as 
    \[ 
        z_i = x_i \oplus y_i \oplus c_{i-1} \textit{ } \forall i \in [1, \ell] \,, 
    \] 
    where $z_i$ denotes the $i$-th bit of the arithmetic sum and $c_i$ denotes the carry bit, such that 
    \[ c_{i} = \left((x_i \oplus c_{i-1}) \land (y_i \oplus c_{i-1})\right) \oplus c_{i-1} \textit{ } \forall i \in [1,\ell] \,, \]
    given $c_0 = 0$. 
\end{definition}

In the conversion from arithmetic secret sharing to boolean, detailed in Algorithm~\ref{alg:design:a2b}, the parties locally define the individual arithmetic shares as boolean, add these using the RCA circuit and obtain boolean shares of the secret.

\begin{algorithm}
    \centering
    \caption{Arithmetic to Boolean share conversion $\Pi_\mathsf{A2B}$}
    \label{alg:design:a2b}
    \begin{algorithmic}[1]
        \Require Party index $i$, $\rssa{x}$
        \Ensure $\rssb{x}$
            \State\label{step:a2b:convert}$P_1$ sets $\rssb{x_1}_1 := (x_1, 0)$, $\rssb{x_2}_1 := (0, x_2)$ and $\rssb{x_3}_1 := (0,0)$.
            \State$P_2$ sets $\rssb{x_1}_2 := (0, 0)$, $\rssb{x_2}_2 := (x_2, 0)$ and $\rssb{x_3}_2 := (0,x_3)$.
            \State$P_3$ sets $\rssb{x_1}_3 := (0, x_1)$, $\rssb{x_2}_3 := (0, 0)$ and $\rssb{x_3}_3 := (x_3, 0)$.
            \State\label{step:a2b:add}The parties compute $\rssb{x} := \mathsf{RCA}(\mathsf{RCA}(\rssb{x_1}, \rssb{x_2}), \rssb{x_3})$.
    \end{algorithmic}
\end{algorithm}

The conversion from boolean secret shares to arithmetic shares is detailed in Algorithm~\ref{alg:design:b2a}. The parties first obtain pairwise uniformly random values, where $r_1$ is drawn between parties $1$ and $2$ and $r_2$ is drawn between parties $2$ and $3$. The parties then proceed to compute the RCA circuit computing shares of $r_3 = x + r_1 + r_2$. Note that the arithmetic replicated secret shares can then simply be derived by revealing $r_3$ to parties $1$ and $3$, and defining the shares to be $(r_3, -r_1, -r_2)$.

\begin{algorithm}
    \centering
    \caption{Boolean to arithmetic share conversion $\Pi_\mathsf{B2A}$}
    \label{alg:design:b2a}
    \begin{algorithmic}[1]
        \Require Party index $i$, $\rssb{x}$
        \Ensure $\rssa{x}$
            \State\label{step:b2a:r1}If $i \neq 3$: $r_1 \xleftarrow{\$_{1,2}} \mathbb{Z}_{2^\ell}$ and $P_1$ sets $\rssb{r_1}_1 := (r_1, 0)$ and $P_2$ sets $\rssb{r_1}_2 := (0, r_1)$.
            \State$P_3$ sets $\rssb{r_1}_3 := (0,0)$.
            \State\label{step:b2a:r2}If $i \neq 1$: $r_2 \xleftarrow{\$_{2,3}} \mathbb{Z}_{2^\ell}$ and $P_2$ sets $\rssb{r_2}_2 := (r_2, 0)$ and $P_3$ sets $\rssb{r_2}_3 := (0, r_2)$.
            \State$P_1$ sets $\rssb{r_2}_1 := (0,0)$.
            \State\label{step:b2a:add}Compute $\rssb{r_3} = \mathsf{RCA}(\mathsf{RCA}(\rssb{x},\rssb{r_1}), \rssb{r_2})$
            \State\label{step:b2a:reveal}$\mathsf{Reveal}$ $r_3$ to $P_1$ and $P_3$
            \State$P_1$ sets $\rssa{x}_1 := (r_3, -r_1)$, $P_2$ sets $\rssa{x}_2 := (-r_1, -r_2)$ and $P_3$ sets $\rssa{x}_3 := (-r_2, r_3)$.
    \end{algorithmic}
\end{algorithm}

\paragraph{Encoding Fixed-Point Arithmetic}
Due to the complexity of ML operations and the high multiplicative depth of the forward pass in a neural network, and to minimize the computation cost, secure ML protocols are designed over rings, more precisely rings of $32$ or $64$-bit integers leveraging the natural CPU types. For this reason, the state-of-the-art frameworks that focus on secure inference work with fixed-point arithmetic. To encode a decimal real number $\Tilde{x} \in \mathbb{R}$ in a fixed-point integer representation $x \in \mathbb{Z}_{2^\ell}$, define $x = \lfloor \Tilde{x}\cdot 2^f \rfloor$, where $f$ denotes the fixed-point precision. After a multiplication, the precision is doubled and truncation is needed to restore the original precision~\cite{zbudila_master_2025}.

\subsection{Fully Homomorphic Encryption}
Fully homomorphic encryption schemes are cryptosystems that
allow arbitrary computation to be performed over encrypted data by any party, independently of whether such a party is able to decrypt ciphertexts. Historically, several schemes provided homomorphic operations for a single operation, such as RSA \cite{RSA} with regards to multiplication or the Pailler cryptosystem~\cite{Paillier1999} with regards to addition. In 2009, the first \emph{fully} homomorphic encryption (FHE) scheme was proposed by Gentry based on ideal lattices~\cite{Gentry2009}. The current state-of-the-art consists of two types of FHE schemes, accumulator-based and SIMD-type schemes.

Accumulator-based schemes such as TFHE~\cite{TFHE} or FHEW~\cite{FHEW} operate on ciphertexts encrypting small plaintext space ranging from 4 to 6 bits, but provide a built-in mechanism to efficiently evaluate arbitrary lookup tables. From a theoretical point of view, it is feasible to work with a larger plaintext space but at a significantly higher computational cost. In the case of Single-Instruction Multiple-Data (SIMD) FHE schemes like BGV, BFV, or CKKS~\cite{BGV,BFV1,BFV2,CKKS}, ciphertexts encrypt entire vectors of values and enable efficient, coefficient-wise addition and multiplication over the slots of the vector. Furthermore, the magnitude of entries within any vector is significantly larger than those used in accumulator-based schemes. In general, the latency per operation in SIMD schemes is greater than in accumulator-based schemes, but by considering the cost amortized over the size of the encrypted vector, SIMD schemes may be superior.

In this work, we rely on a SIMD scheme, namely CKKS~\cite{CKKS}, which is introduced in the subsequent paragraph. Note that, by default, FHE schemes only serve to protect the underlying data of a computation but not the computation itself. More specifically, any user obtaining the (encrypted) result of the computation may be able to derive potentially confidential information about the nature of the computation. This may be  detrimental to a model provider aiming to hide certain characteristics of a machine learning model.

\subsubsection{The CKKS Scheme}\label{sec:background:ckks}

The CKKS scheme was introduced by Cheon et al.~\cite{CKKS} and, similarly to other SIMD schemes, encrypts entire vectors. Here, we give a brief overview of operations the scheme provides, but do not discuss the intricacies of the scheme as they are beyond the scope of this article. Let ($\pk, \sk$) denote a public and private key of the CKKS scheme respectively. The cryptosystem provides the following operations.

\begin{itemize}
    \item $\ct \gets \FHEEnc(\vec{m}, \pk)$: The procedure $\FHEEnc$ takes as input a vector
    $\vec{m} \in \RR^{L}$ and a public key $\pk$, and produces a ciphertext $\ct$.
    \item $\vec{m}' \gets \FHEDec(\ct, \sk)$: The decryption procedure $\FHEDec$, upon the input of a ciphertext
    $\ct$ and the secret key $\sk$, outputs a vector $\vec{m}' \in \RR^{L}$. We stress that in the case of the CKKS scheme it may be true that 
    \[\FHEDec(\FHEEnc(\vec{m}, \pk), \sk) = \vec{m} + \vec{\epsilon}\,.\]
    In other words, the decryption procedure may not output the exact
    value we encrypted. Often, the error $\vec{\epsilon}$ is 
    compared to the error incurred due to floating point operations if we operated over the plaintext vector $\vec{m}$ without encryption. If managed properly, this error will not affect the correctness of the computation.
    \item $\ct_{+} \gets \FHEAdd(\ct_0, \ct_1)$: 
    Given $\ct_i = \FHEEnc(\vec{m}_i, \pk), i \in \{0,1\}$,
    the procedure $\FHEAdd$ outputs a new ciphertext
    $\ct_{+}$ such that \[\FHEDec(\ct_{+}) = \vec{m}_0 + \vec{m}_1 + \vec{\epsilon}\,.\]
    \item $\ct_{\odot} \gets \FHEMul(\ct_0, \ct_1)$:
    Given $\ct_i = \FHEEnc(\vec{m}_i, \pk), i \in \{0,1\}$,
    the procedure $\FHEMul$ outputs a new ciphertext
    $\ct_{\odot}$ such that \[\FHEDec(\ct_{\odot}) = \vec{m}_0 \odot \vec{m}_1 + \vec{\epsilon}\,,\] where $\odot$ denotes the Hadamard product.
    \item $\ct' \gets \FHERot(\ct, r), r \in \ZZ$: Given a ciphertext
    $\ct = \FHEEnc(\vec{m}, \pk)$ and a scalar $r$, the procedure $\FHERot$ outputs a ciphertext
    $\ct'$ such that $\FHEDec(\ct', \sk) = \vec{m}' + \vec{\epsilon}$ and $\vec{m}'_i = \vec{m}_{i + r \pmod{L}}$ that is, the procedure outputs a new vector containing the \textit{cyclically shifted} coefficients of $\vec{m}$.
\end{itemize}

Note that the CKKS scheme relies on additional key material beyond a secret key $\sk$ and public key $\pk$, which we suppressed in the notation above. This requirement is not specific to the CKKS scheme but any FHE scheme, though the nature of the key material may vary. Specifically, the operations $\FHEMul$ and $\FHERot$ require so-called relinearization (multiplication) and automorphism (rotation) keys, respectively. Furthermore, the additional key material is intended to be made public and does not pose a security risk.

\subsubsection{Arbitrary Function Evaluation in CKKS}

The CKKS scheme is particularly suited to perform polynomial
operations. However, many functions commonly used in machine learning cannot be exactly represented using polynomials. Hence, we must rely on polynomial approximations that we construct using polynomial interpolation.

Polynomial interpolation is a fundamental technique in numerical analysis and approximation theory, aimed at constructing a function that passes through a given set of data points. The goal is to approximate arbitrary functions using a set of basis functions that are easier to evaluate. Among the various methods of interpolation, polynomial interpolation is widely used due to its simplicity and efficiency and a common choice in FHE-based settings as polynomials can be easily evaluated. However, the choice of interpolation nodes, that is the set of points at which the function values are known, can significantly impact the accuracy and stability of the interpolation. 

We give a general procedure to interpolate a function $f$ using a polynomial. Given a set of \( n+1 \) data points $(x_i, f_i), i=0,1,...,n$ such that $f_i = f(x_i)$, we aim to find a polynomial \( P(x) = \sum_{i = 0}^{n} a_i x^i \) such that: \[ P(x_i) = f_i \quad \text{for} \quad i = 0, 1, \ldots, n\,. \]
The coefficients \( a_0, a_1, \ldots, a_n \) can be determined by solving the system of equations:
\[
\begin{pmatrix}
1 & x_0 & x_0^2 & \cdots & x_0^n \\
1 & x_1 & x_1^2 & \cdots & x_1^n \\
\vdots & \vdots & \vdots & \ddots & \vdots \\
1 & x_n & x_n^2 & \cdots & x_n^n
\end{pmatrix}
\begin{pmatrix}
a_0 \\
a_1 \\
\vdots \\
a_n
\end{pmatrix}
=
\begin{pmatrix}
f_0 \\
f_1 \\
\vdots \\
f_n
\end{pmatrix}\,.
\]
The system matrix is commonly referred to as a Vandermonde Matrix. We stress that the procedure above is unsuitable for many settings as the matrix may be poorly conditioned. 

In cases for which we may pick the set of nodes, it is possible to enforce certain properties of the interpolant. Chebyshev nodes are specific points used in interpolation to minimize the maximum interpolation error over an interval. They are defined as the roots of the Chebyshev polynomials of the first kind, \( T_{n+1}(x) \), which are defined on the interval \([-1, 1]\). The Chebyshev nodes \( x_k \) are given by:

\[ x_i = \cos\left(\frac{2i + 1}{2n + 2} \pi\right) \quad \text{for} \quad i = 0, 1, \ldots, n\,. \]
These nodes are particularly useful because they help reduce the Runge phenomenon, characterized by large oscillations at the edges of the interval when using equally spaced points. Furthermore, by leveraging Chebyshev nodes it is possible to bound the absolute interpolation error  for any $x \in [-1, 1]$ as

\[|f(x) - P(x)| \leq \frac{1}{2^n (n+1)!} \max_{\bar{x} \in [-1,1] }|f^{(n+1)}(\bar{x})|\,, \] where $f^{(n+1)}$ denotes the $n+1$-th derivative of $f$. Finally, we note that extending the procedure to arbitrary intervals instead of $[-1,1]$ can be achieved through a change of variables, and that Chebyshev nodes often perform well even if $f$ is not differentiable.

\section{Platform Design and Implementation Details}\label{sec:design}

In the following section, we detail the general design and implementation details of the MOZAIK platform. We start by outlining the design goals of the system, which serve as the foundation for our design and development process.

\subsection{Design Goals}\label{sec:design-goals}
The MOZAIK project aims to design a novel system architecture that improves security, preserves privacy, and enables scalable, end-to-end secure computation and analytics on sensitive IoT data within a cloud-assisted environment.
Such a secure, scalable, and privacy-preserving computation platform must achieve the following key design goals.

\begin{description}   
    \item[End-to-End Data Confidentiality.] Ensure that IoT data is encrypted at the source before it leaves the user's domain, and that it remains encrypted throughout transmission, storage, and processing, without exposing sensitive information to any external or cloud-based entity.

    \item[User Control and Trust.] Establish a user-centric trust model that places the user in full control of their data throughout its lifecycle. Users should be required to give their explicit consent to data processing on specific subsets of their data and only by data processors approved by the user.
    
    \item[Correctness.] The accuracy of all data processing and analytics operations in the system should match the results that would be obtained if the same computations were performed directly on the plaintext data without any privacy-preserving mechanism in place.
    
    \item[Scalability.] Support large-scale IoT deployments with a system that is capable of handling large volumes of data generated by numerous distributed devices, while maintaining performance and low latency suitable for applications that require real-time or near real-time processing.
    
    \item[Interoperability.] Provide a modular and extensible system that can easily integrate with various IoT devices, cloud providers, and data processing entities, promoting broad applicability across different use cases.
    
    \item[Resilience Against Threats.] Ensure that the system is robust against a wide range of attack vectors, including unauthorized access, data leakage, malicious cloud service providers, and adversarial attacks on encrypted computation.

\end{description}

\subsection{MOZAIK Platform Design Overview}\label{sec:design-mozaik}

\begin{figure}
    \centering
    \includegraphics[width=1.0\linewidth]{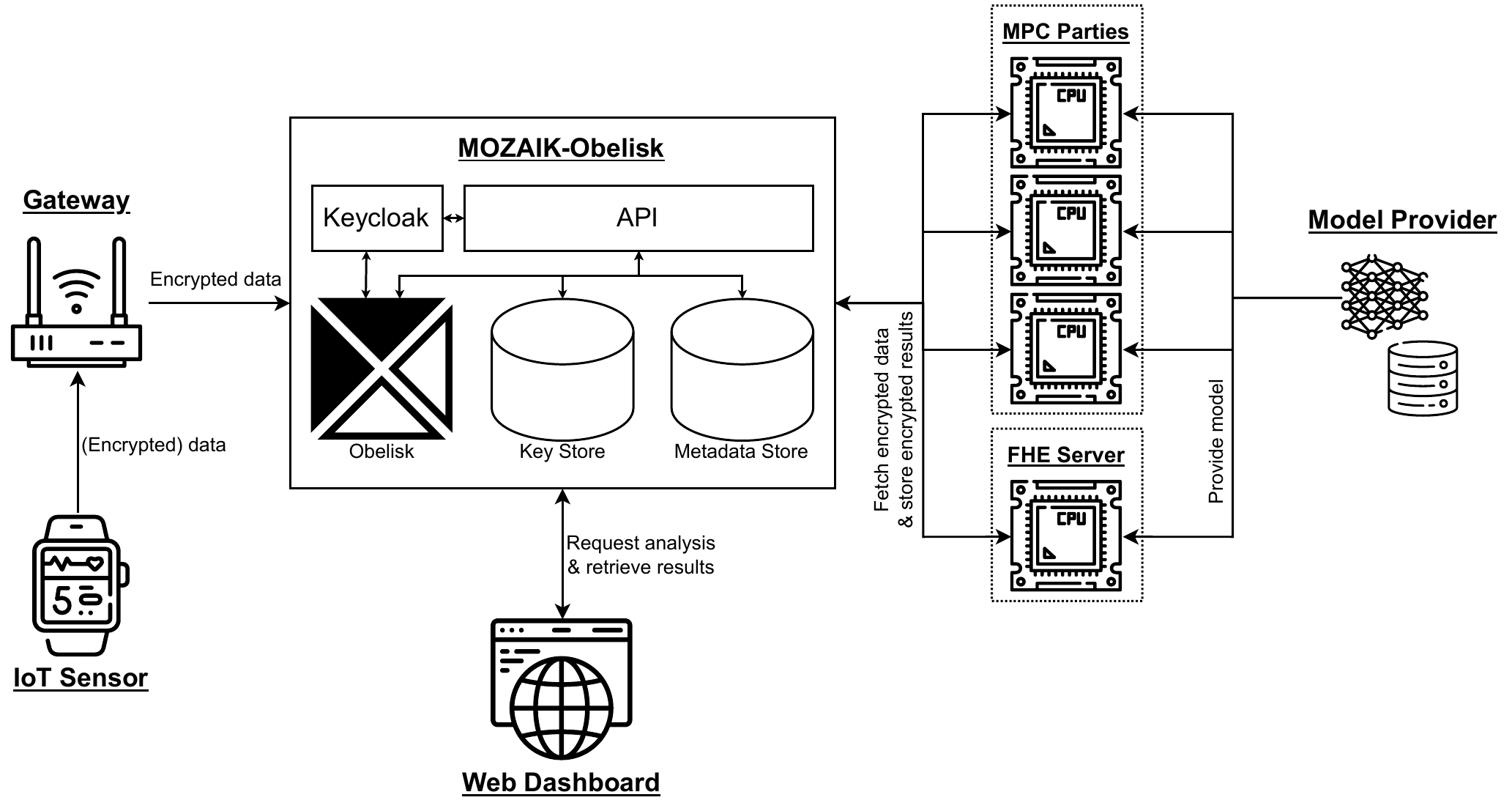}
    \caption{Schematic overview of the proposed proof-of-concept system architecture of MOZAIK. The data generated by an IoT device is encrypted and securely stored in Obelisk. The FHE server or MPC parties perform computation on the encrypted data on request while ensuring user privacy and store the encrypted result back into Obelisk. The model provider supplies a machine learning model to the computation servers for inference on the user's data. This figure contains resources from Flaticon.com.}\label{fig:design:mozaik_architecture}
\end{figure}

Based on the aforementioned design goals, we propose a proof-of-concept (PoC) system architecture to realize secure and scalable privacy-preserving IoT data processing. The PoC implementation is composed of several core components, depicted in Figure~\ref{fig:design:mozaik_architecture}.
These components are IoT sensors, a gateway, the MOZAIK-Obelisk data storage platform, a web dashboard, data processing servers (MPC parties or an FHE server), and a model provider.

Users are in control of one or more IoT devices that produce or collect personal or sensitive data using sensors. Although not a requirement, a user can also be in control of a gateway, which acts as a single data collection point for multiple IoT devices from that user. The gateway can support other connectivity technologies like Bluetooth to directly communicate with and collect data generated by sensors, allowing support for a wider range of IoT devices or sensor types. Collected data from IoT devices is encrypted before leaving the user's control. This can occur on the IoT device itself, or on the gateway, depending on the user's preference, the specific data collection scenario or location, or because of IoT device limitations. The encrypted data is then sent to the platform for storage.

All aspects of data storage within the MOZAIK system are unified under what we call MOZAIK-Obelisk. It contains a stateless, high-level application programming interface (API) that provides an abstraction layer to the underlying components of MOZAIK-Obelisk, enhancing modularity and interoperability. The encrypted IoT data itself is stored in Obelisk~\cite{bracke_design_2021}, an independently scalable data platform designed for IoT applications (as discussed in Section~\ref{sec:background:obelisk}), waiting to be processed.

When a computation is triggered, either manually by a user via the web dashboard or automatically when streamed processing is enabled, the API calls a standardized endpoint on the chosen MPC parties or FHE server to request the computation. The computation servers then obtain a machine learning model from a model provider, fetch the computation-specific encrypted data from Obelisk, and perform the requested computation. The encrypted computation results are subsequently stored in Obelisk and made available to the data owner, who is the only one that has the correct keys to decrypt the result.


\subsection{Threat Model}\label{sec:design:threat-model}
The threat model details our assumptions about the adversarial behavior of the different entities in the system.
It is an important part to appropriately model security goals in the system and to validate our employed protection techniques.

We consider two types of adversarial behavior. \emph{Semi-honest} (or passive) adversaries follow the protocol honestly, nevertheless try to gain information from the data available in the process. On the other hand, we say an adversary is \emph{malicious} (or active) if it deviates arbitrarily from the protocol in order to violate security.

Our threat model is as follows.
We assume that a semi-honest/malicious adversary $\mathcal{A}$ fully controls multiple users, the central storage platform (MOZAIK-Obelisk), and controls the processing entity to the degree described below in Sections~\ref{sec:design:threat-model:mpc} and~\ref{sec:design:threat-model:fhe} (depending on the COED technology used). The goal of $\mathcal{A}$ is to violate any of the following design goals (defined in Section~\ref{sec:design-goals}) involving another user: user control, user trust, privacy, or correctness.
We also allow $\mathcal{A}$ to control the model provider \emph{semi-honestly} to support its goal. Specifically, we aim to model a setting where the model provider does not learn any information about the user's sensitive data. At the same time, we want to exclude a variety of possible attacks where the model provider could selectively change the model parameters or architecture to gain insights from user inputs. Since our work is about the processing part, we consider these attack vectors out of scope and therefore only allow semi-honest corruption of the model provider.

\emph{Note.} The goal of our adversary is to attack the user of the platform, not the model (provider). While we allow the adversary to use other malicious users to gain an advantage, we do not want to claim full privacy for the model parameters. Attacks to (partially) learn the model parameters or training data, such as model inversion or membership inference attacks, are out of scope.

\subsubsection{Threat Model for MPC-based Processing}\label{sec:design:threat-model:mpc}
When MPC is used to perform data processing, we assume that in addition to corrupting multiple users, the central storage platform, and the model provider, the adversary also corrupts at most $t$ out of $n$ MPC parties. Concretely, in our PoC architecture, we use three MPC parties and allow up to one semi-honest or malicious corruption.

\subsubsection{Threat Model for FHE-based Processing}\label{sec:design:threat-model:fhe}
If FHE is used for data processing, we assume that the adversary additionally has semi-honest access to the FHE computation server.

By construction of an FHE scheme, any malicious adversary is capable of manipulating ciphertexts. Therefore, an adversary can
target a specific user by strategically causing computations to fail and may incrementally obtain information about the encrypted data. Thwarting such attacks is possible by relying on zero-knowledge proofs, but may cause a significant overhead in practice. For additional details on this we refer the reader to~\cite{HELIOS,blind-zksnark}. Finally, we stress that publishing intermediate or final results of an FHE computation may create security risks if the primitives are used improperly; see~\cite{CPA1,CPA2}.

\subsection{MOZAIK Component Implementation Details}\label{sec:design:implementation-details-mozaik} 
We now turn to describing each system component separately.

\subsubsection{Data Processing Modes}\label{sec:design:details:data-processing-modes}

\begin{figure}
    \centering
    \begin{subfigure}[b]{0.49\linewidth}
        \centering
        \includegraphics[width=0.95\linewidth]{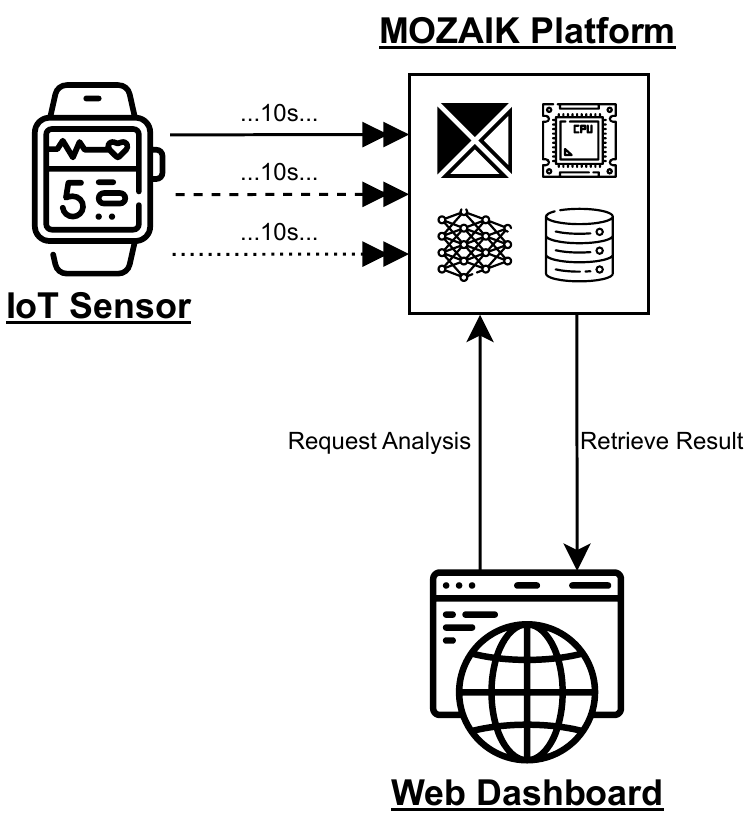}
        \caption{Ad Hoc Analysis Mode}
    \end{subfigure}
    \hfil
    \centering
    \begin{subfigure}[b]{0.49\linewidth}
        \centering
        \includegraphics[width=0.95\linewidth]{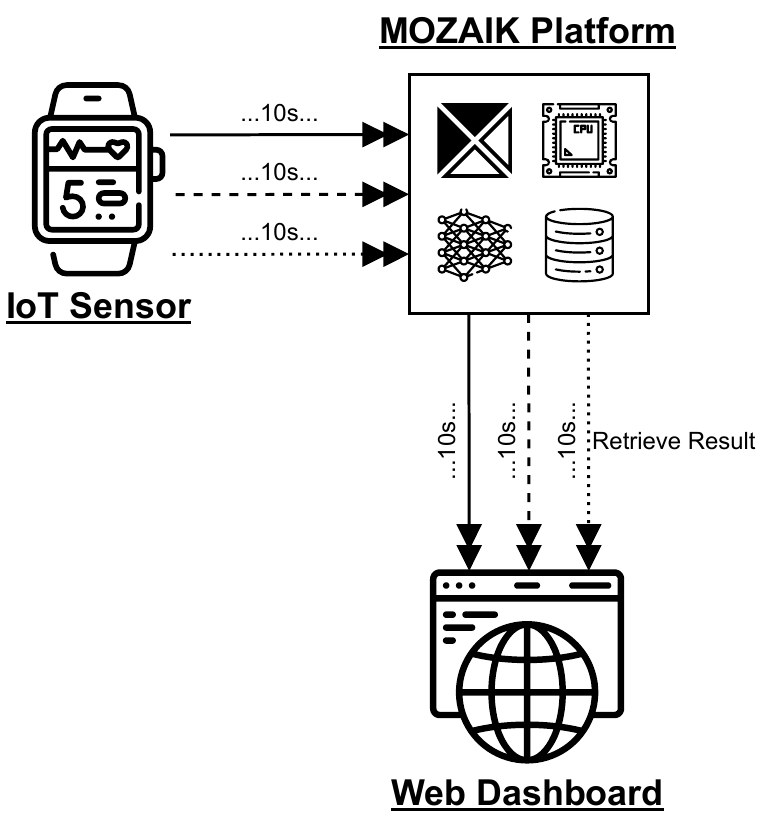}
        \caption{Streaming Mode}
    \end{subfigure}
    \caption{Visualization of the ad hoc analysis (a) and streaming (b) mode provided by the MOZAIK platform. In ad hoc analysis mode, ingested data is only processed upon explicit request by the user. In streaming mode, the user explicitly grants the continuous processing of their data for a certain time period. This figure contains resources from Flaticon.com.}\label{fig:design:processingmodes}
\end{figure}

IoT sensor devices typically generate data or capture measurements at a certain frequency. This data is then ingested into the platform at the same rate, with an added timestamp to ensure it can be efficiently searched and indexed. As data is continuously ingested, two distinct data processing modes emerge: ad hoc analysis and streaming. Both modes are detailed in the following and conceptually depicted in Figure~\ref{fig:design:processingmodes}.

\paragraph{Ad Hoc Analysis}
In ad hoc analysis mode, the user has full control over when data is processed. No data point is analyzed without the user's explicit permission or involvement. The user selects which data to process by specifying a time range and grants explicit consent to perform the analysis. Multiple data points can be processed in parallel by grouping them in a batch. The analysis results of the selected data points are then delivered to the user within the expected analysis latency.

\paragraph{Streaming}
In streaming mode, the user activates streaming for an IoT sensor device once for a specified amount of time, during which they provide consent for continuous automated data processing without further explicit approval for the specified period. The user can also opt to stop streaming at any point before the specified time ends. Once streaming is activated, the platform groups the ingested data points in micro-batches and analyzes each batch as soon as possible, providing the user with near real-time data insights. It is essential to keep the streaming ingestion rate lower than the end-to-end computation latency, otherwise the system may become overloaded.

\subsubsection{Sensor Data Encryption Details}\label{sec:design:details:encryption}
We distinguish two locations where the collected IoT sensor data is encrypted: on the IoT device itself or on a gateway. The choice of location depends on the capabilities and resource constraints of the IoT device, the specific data collection scenario, the availability of a gateway, and the adversarial model.

\paragraph{Encryption on the IoT Sensor Device}
If the IoT sensor device is capable of computing symmetric authenticated encryption, then encryption directly on the device is the simplest option.

Since nonce-based schemes require a fresh nonce for every encryption, we require the device to hold a small state $\st$ across sensing operations, from which we derive a fresh nonce $N$\footnote{For example, this can be a counter which is incremented every time a new nonce is generated. Once the counter overflows the nonce domain, we need to rotate the key $k$.}.
Let $d$ be the sensor data point, the encryption is computed as
\begin{equation}\label{eq:design:enc-on-device}
    \ctsymm_d^k := \Enc_k(d, N, \mathsf{ID}_\mathsf{user} \| N) \,,
\end{equation}
where $\mathsf{ID}_\mathsf{user}$ is the user ID of the owner of the IoT sensor device.
Then, $N$ and $\ctsymm_d^k$ are sent to MOZAIK-Obelisk.

\paragraph{Encryption on the Gateway}
In case the IoT sensor device is not capable of performing the encryption itself, or if data from multiple sensor devices is aggregated through a gateway, the encryption process can be delegated to the gateway.
We do not specify by which method data is transmitted from the sensor device(s) to the gateway, but we assume this transmission occurs securely within the user's control domain. At the gateway, a nonce $N$ is sampled, the corresponding device key $k$ is retrieved or a gateway-specific key is used, and the data point is encrypted following the same procedure as in Equation~\ref{eq:design:enc-on-device}.

\subsubsection{Central Data Storage: MOZAIK-Obelisk}\label{sec:design:details:mozaik_obelisk}

One of MOZAIK's goals is to securely store sensitive (encrypted) data, even if only temporarily, in order to perform privacy-preserving computation and provide users with insights into their data. As mentioned above, we utilize Obelisk HFS (see Section~\ref{sec:background:obelisk}) to store the encrypted IoT sensor data. Completing the entire end-to-end data flow, from the IoT sensor device to the computation server(s) and ultimately to reporting the analysis result to the user, involves numerous components that must interact reliably and efficiently with each other. To address this, an overarching API is designed that facilitates seamless communication between all components and entities within the system. It serves as the single external access point for communicating with any component within MOZAIK-Obelisk.

The overarching API is designed to be entirely stateless, meaning that it does not retain any information in memory between requests. Each request is treated completely independently and self-contained, with no reliance on previous interactions. This stateless design eliminates the need for the API to track client state, facilitating horizontal scaling. As a result, performance and reliability can be improved through load balancing and failover across multiple API instances.

The API provides a high-level abstraction of the underlying components of MOZAIK-Obelisk, creating a modular and interoperable system that reduces the dependency on any particular software vendor. For example, if a specific use case demands a different data store, Obelisk can be replaced by another data store without affecting the interface with IoT devices, users, and computation servers. In such case, only the abstraction layer would need to be modified to support the replacement data store.

To enhance security, the API performs additional authentication and authorization checks using Keycloak and JSON Web Token (JWT)~\cite{rfc7519,rfc8725}. It acts as the initial safeguard against unauthorized data access. For instance, when a user seeks to perform a computation on a specific data point within a dataset, the API verifies whether the user has the necessary permissions to access that dataset. The API will delegate the computation request to the designated computation server(s) to initiate the analysis process if and only if the authorization was successful.

The key store and metadata store are two separate Redis~\cite{redis_redis_nodate} instances. The key store temporarily stores encrypted key shares (described below in Section~\ref{sec:design:details:key-management}), while the metadata store maintains information about each requested analysis, including a unique analysis ID, the Obelisk dataset ID for sensor data, the target dataset ID for storing results, the participating computation servers, and other related metadata. Obelisk, the key store, and the metadata store interact solely with the overarching API and not with each other directly. Additionally, Kubernetes network policies~\cite{kubernetes_network_policies_nodate} enforce this isolation, permitting communication only with the API (and Keycloak in Obelisk's case), to mitigate lateral movement attacks.

\subsubsection{Encryption Key (Share) Management Details}\label{sec:design:details:key-management}

\paragraph{Key Management for FHE}
We start by describing key management for FHE, as it is conceptually simpler than that of MPC. When an FHE-based analysis is initiated, the request is sent to a compute engine, which then retrieves the associated ciphertext and the necessary auxiliary key material linked to the user ID from Obelisk. Since the ciphertexts are encrypted under the FHE scheme, data processing can begin immediately, without requiring any additional information apart from the ciphertext and auxiliary key material.

\paragraph{Key (Share) Management for MPC}
For MPC-based processing, IoT sensor data is encrypted using a symmetric encryption scheme with the user's or device's secret key $k$. Before data analysis can begin, the MPC processing parties must receive sufficient information about $k$ in the form of additive secret shares of $k$.

Since only the user knows this secret key, the key shares $k_i$ are created locally by the user using a $\mathsf{SecretShare}$ function such that $k_1 \oplus k_2 \oplus k_3 = k$. These shares should then be included in the analysis request to the overarching API. However, in the MPC setting, it is crucial that the central data storage does not gain direct access to these key shares, as possessing all of them would allow an adversary who compromises the central data storage to reconstruct the key and breach data privacy.

Instead, we employ the key management protocol of Marquet et al.~\cite{marquet_secure_2023}. Here, the user selects the MPC parties that will perform the computation beforehand and encrypts each key share with the corresponding party's public key, i.e., $\ctsymm_{k_i}^{\pk_{i}} := \Enc_{\pk_i}(k_i, \ad_i)$, where $\pk_{i}$ is the public key of party $i$ and $\ad_i$ denotes associated data for this encryption.
So, instead of sending the key shares directly to the overarching API, only $\ctsymm_{k_1}^{\pk_{1}}, \ctsymm_{k_2}^{\pk_{2}}$ and $\ctsymm_{k_3}^{\pk_{3}}$ are sent to the API during the analysis request, and these encrypted key shares are temporarily stored in MOZAIK-Obelisk's key store (see Figure~\ref{fig:design:mozaik_architecture}) until the computation request completes or expires. We require that $\Enc$ is an IND-CCA(2) secure public-key encryption scheme, e.g., RSA-OAEP~\cite{C:FOPS01} or ECIES~\cite{AbdBelRog98}.

The associated data needed to encrypt each key share is constructed as follows. Let $\mathsf{ID}_\mathsf{user}$ be a unique user ID, $\mathsf{ID}_\mathsf{data}$ be a set of data point identifiers, e.g., timestamps, $\mathsf{type}$ be the analysis type (i.e., the requested analysis algorithm or inference model), and $\mathsf{alg}$ be the identifier for the symmetric encryption algorithm that the IoT sensor device used to encrypt the sensor data (e.g., AES-128-GCM), we define the associated context data as
\begin{align*}
    \ad_i & := \mathrm{0x01} \| \mathsf{ID}_\mathsf{user} \| \pk_1 \| \pk_2 \| \pk_3 \| \mathsf{ID}_\mathsf{data} \| \mathsf{type} \| \mathsf{alg} \| \pk_i & \text{for ad hoc analysis} \,, \\
    \ad_i & := \mathrm{0x02} \| \mathsf{ID}_\mathsf{user} \| \pk_1 \| \pk_2 \| \pk_3 \| t_{\mathsf{begin}} \| t_{\mathsf{end}} \| \mathsf{type} \| \mathsf{alg} \| \pk_i & \text{for streaming} \,,
\end{align*}

where $t_{\mathsf{begin}}$ and $t_{\mathsf{end}}$ denote the user-defined start and expiration times of the streaming request.

When an MPC party receives a computation request, it also receives additional metadata, such as the user ID, the data indices that require analysis, and the group of MPC parties that will perform the processing. The MPC party derives $\ad_i$ from the received metadata and requests $\ctsymm_{k_i}^{\pk_{i}}$ from the API, after which it attempts to decrypt the ciphertext key share with $\sk_i$ to obtain a key share $k_i$ that will be the input for the data processing step (described in Section~\ref{sec:implementation:data-processing:mpc}). Note that in the case of streaming, the MPC party will also check if the current time is in the user-specified time window for which streaming was allowed, i.e., $t_{\mathsf{begin}} \le t \le t_{\mathsf{end}}$. If not, the MPC party refuses the operation. This prevents that a malicious central storage entity attempts to process data with expired key material.

This approach has a central purpose in our design. The associated data binds the encrypted key shares to a processing context. An honest MPC party cannot be coerced into performing a different analysis, using different data points, involving different MPC parties, or switching between streaming and ad hoc computation, other than specified in the processing parameters selected by the user when the analysis request was created. Therefore, the created ciphertexts $\ctsymm_{k_i}^{\pk_i}$ can be viewed as a technical part of the consent the user gives for data processing. The processing context increases the obstacles for abuse and works towards user trust. Additionally, authenticating the encrypted key share detects malicious modification of the ciphertext by an intermediary party, such as the central data storage, thereby preventing decryption oracle attacks.

\subsubsection{Data Processing}\label{sec:design:details:coed}
The MOZAIK platform supports two methods of processing encrypted data: MPC and FHE. The following overview outlines the general interaction and data flow between MOZAIK-Obelisk and the processing servers for both MPC- and FHE-based processing. Specific implementation details, which are highly use-case dependent, are provided in Section~\ref{sec:implementation:data-processing}.

\begin{figure}
    \centering
    \includegraphics[width=1.0\linewidth]{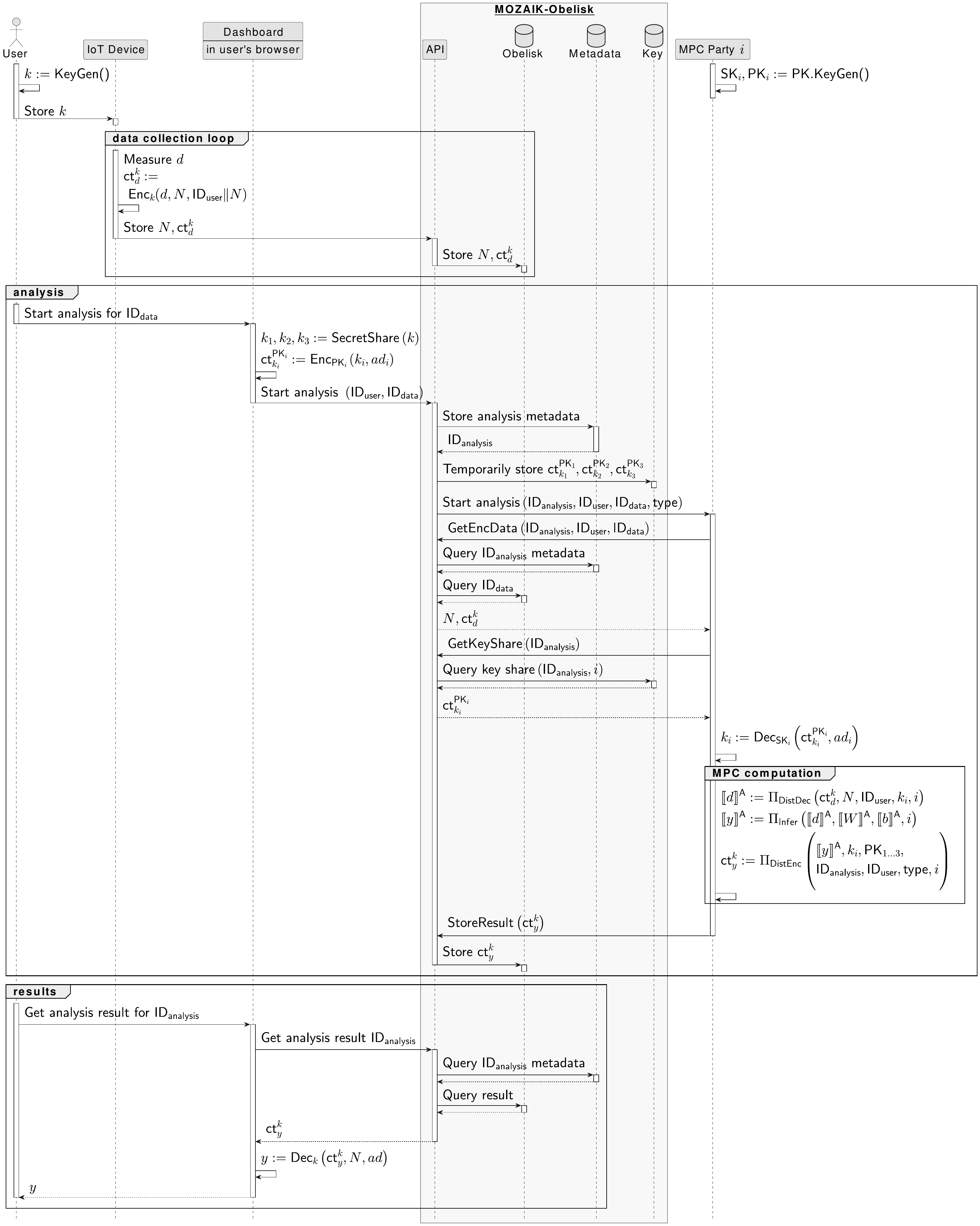}
    \caption{Sequence diagram illustrating the complete end-to-end flow of an ad hoc MPC analysis. An IoT sensor device collects and encrypts data points, after which they are stored within the user's dataset in Obelisk. The user later initiates an ad hoc analysis on selected data points, triggering the computation flow. Finally, the user obtains the analysis result.}\label{fig:implementation:computation_sequence_diagram}
\end{figure}

\paragraph{MPC}\label{sec:implementation:computation-flow:mpc}
MPC programs interact with MOZAIK-Obelisk through a REST API. An overview of the flow of requests, data, and computation for an ad hoc analysis is illustrated in Figure~\ref{fig:implementation:computation_sequence_diagram}.

\subparagraph{Processing}
Algorithm~\ref{alg:mpcprocessing} describes MOZAIK's MPC-based processing at a high level. To initiate a computation, the MOZAIK-Obelisk's API sends a request to each MPC party containing a unique analysis ID ($\mathsf{ID}_\mathsf{analysis}$), the user's unique ID ($\mathsf{ID}_\mathsf{user}$), a set of data point identifiers ($\mathsf{ID}_\mathsf{data}$) and analysis type ($\mathsf{type}$), specifying the analysis to be performed. Each MPC server then adds this request to a queue for sequential processing.
For each request in the queue, the parties proceed with the computation as follows. First, the MPC parties fetch the requested encrypted data points from Obelisk and the corresponding encrypted symmetric key shares from the key store. Note that the data points are encrypted under a symmetric key $k$, whereas each corresponding secret share of $k$ for party $i$ is encrypted with their corresponding public key $\pk_i$.
The MPC party locally decrypts the secret share $k_i$ of the symmetric key $k$ using $\sk_i$ and then jointly runs the MPC protocol for distributed decryption among the other MPC parties indicated in the request.
The obtained shares of each data point are used in the subsequent MPC protocol to compute the desired inference task. Note that we assume the machine learning model to be evaluated had been pre-shared by the model provider prior to this service. As a result, parties store their shares of the model locally and retrieve them for computing the inference.
Finally, the resulting share is encrypted via a distributed encryption protocol, and the encrypted result is sent back to MOZAIK-Obelisk where it is stored in Obelisk in the user's result dataset.

\begin{algorithm}
\caption{MPC-based data processing}
\label{alg:mpcprocessing}
    \begin{algorithmic}[1]
    \Require Party index $i$, private-public key pair $\left(\sk_i,\pk_i\right)$, shares of a machine learning model $\left(\rssa{W}, \rssa{b}\right)$
    \State\label{state:mpcprocessing:queue}$\mathsf{queue} \mathrel{{+}{=}} (\mathsf{ID}_\mathsf{analysis}, \mathsf{ID}_\mathsf{user}, \mathsf{ID}_\mathsf{data}, \mathsf{type}) \xleftarrow[\mathsf{HTTPS}]{} \mathsf{MozaikObelisk}$
    \For{$\mathsf{request}$ \textbf{in} $\mathsf{queue}$}
        \State\label{state:mpcprocessing:getdata} $N,\ctsymm_d^k  \xleftarrow[\mathsf{HTTPS}]{\mathsf{GetEncData}(\mathsf{ID}_\mathsf{analysis}, \mathsf{ID}_\mathsf{user}, \mathsf{ID}_\mathsf{data})} \mathsf{MozaikObelisk}$
        \State\label{state:mpcprocessing:getkeyshare} $ \ctsymm^{\pk_i}_{k_i} \xleftarrow[\mathsf{HTTPS}]{\mathsf{GetKeyShare}(\mathsf{ID}_\mathsf{analysis})} \mathsf{MozaikObelisk}$
        \State\label{state:mpcprocessing:pkdec} $ k_i := \mathsf{Dec}_{\sk_i}\left(\ctsymm^{\pk_i}_{k_i},ad_i\right)$
        \State\label{state:mpcprocessing:distdec} $\rssa{d} := \Pi_{\mathsf{DistDec}}\left(\ctsymm^k_d, N, \mathsf{ID}_\mathsf{user}, k_i, i\right)$ \hfill (see Algorithm~\ref{alg:design:dist-dec})
        \State\label{state:mpcprocessing:infer} $\rssa{y} := \Pi_{\mathsf{Infer}}\left(\rssa{d}, \rssa{W}, \rssa{b}, i\right)$ \hfill (see Algorithm~\ref{alg:design:infer})
        \State\label{state:mpcprocessing:distenc} $\ctsymm^k_y := \Pi_{\mathsf{DistEnc}}\left(\rssa{y}, k_i, \pk_{1\ldots{}3}, \mathsf{ID}_\mathsf{analysis}, \mathsf{ID}_\mathsf{user}, \mathsf{type}, i\right)$ \hfill (see Algorithm~\ref{alg:design:dist-enc})
        \State\label{state:mpcprocessing:store} $\ctsymm^k_y \xrightarrow[\mathsf{HTTPS}]{\mathsf{StoreResult}} \mathsf{MozaikObelisk}$
    \EndFor
    \end{algorithmic}
\end{algorithm}

\subparagraph{Batched Processing}
Batched processing is similar to the processing of a single analysis, except that it operates on all data points from multiple analysis requests simultaneously, improving processing throughput. Note that, for simplicity, Algorithm~\ref{alg:mpcprocessing} does not inherently support batched requests with multiple users. In order to support batched computation for our architecture, MOZAIK-Obelisk aggregates computation requests from multiple users in a batch and sends them collectively to the MPC parties. As a result, the MPC parties receive a vectorized version of the previously described variables (i.e., a vector of $\mathsf{ID}_\mathsf{analysis}$, $\mathsf{ID}_\mathsf{user}$, and $\mathsf{ID}_\mathsf{data}$), which is naturally supported by the underlying MPC protocols. The distributed decryption protocol is run on a vector of encrypted data, with a list of symmetric key shares $[k_i]$ as input, such that the $i$-th key share corresponds to the $i$-th requested analysis. The output of the distributed decryption is in the form of a vector of matrices, where the rows of each $i$-th matrix correspond to the selected data points (${\mathsf{ID}_\mathsf{data}}_i$) from the $i$-th analysis. For inference, this structure is flattened into a single large matrix, with each row corresponding to an individual data point. The parties then perform ML inference on all the provided data at once. The inference output is then unflattened into a vector of matrices, where the rows of the $i$-th matrix represent the prediction confidence levels for data points from the $i$-th analysis. The parties then run the distributed encryption protocol, encrypting each $i$-th matrix with the corresponding $i$-th key. Finally, the parties store the ciphertexts in Obelisk.

\paragraph{FHE}\label{sec:implementation:computation-flow:fhe}
At a high level, the FHE-based processing is designed in a way similar 
to the MPC-based processing and also relies on the REST API for communication with MOZAIK-Obelisk.

FHE-based processing is described in Algorithm~\ref{alg:fheprocessing}. Each time a user wishes to perform an analysis, a request is sent to an FHE-Engine by the API which is then added to a task queue. For every request in the queue, the relevant ciphertext is fetched from Obelisk. Then, the required key material needs to be loaded onto the local storage of the FHE-Engine. In practice, the size of the material may be considerable (40 MiB--200 GiB), but is associated with a specific user rather than a particular computation. Hence, we cache the key material locally on the FHE-Engine and only load the material if it is not present yet. Finally, the inference is performed using Algorithm~\ref{alg:fheinf} and the result is sent back to MOZAIK-Obelisk.

\begin{algorithm}
\caption{FHE-based data processing }
\label{alg:fheprocessing}
\begin{algorithmic}[1]
\State\label{state:fheprocessing:queue}$\mathsf{queue} \mathrel{{+}{=}} (\mathsf{ID}_\mathsf{analysis}, \mathsf{ID}_\mathsf{user}, \mathsf{ID}_\mathsf{data}) \xleftarrow[\mathsf{HTTPS}]{} \mathsf{MozaikObelisk}$

\For{$\mathsf{request}$ \textbf{in} $\mathsf{queue}$}
    \State $\ct \xleftarrow[\mathsf{HTTPS}]{\mathsf{GetEncData}(\mathsf{ID}_\mathsf{analysis}, \mathsf{ID}_\mathsf{user}, \mathsf{ID}_\mathsf{data})} \mathsf{MozaikObelisk}$\label{state:fheprocessing:getdata}
    \If{$\mathsf{key\_cache}[\mathsf{ID}_{\mathsf{user}}] = \varnothing$}
        \State\label{state:fheprocessing:keys} $\mathsf{KM} \xleftarrow[\mathsf{HTTPS}]{\mathsf{GetKeyMaterial}(\mathsf{ID}_\mathsf{user})}\mathsf{MozaikObelisk}$
        \State\label{state:fheprocessing:keystore} $\mathsf{key\_cache}[\mathsf{ID}_\mathsf{user}] := \mathsf{KM}$
    \Else 
        \State\label{state:fheprocessing:load} $\mathsf{KM} := \mathsf{key\_cache}[\mathsf{ID}_\mathsf{user}]$
    \EndIf 
    \State $\ct' := \mathsf{FHEInfer}(\ct,\mathsf{KM})$ \hfill (see Algorithm~\ref{alg:fheinf})\label{state:fheprocessing:infer} 
    \State $\ct' \xrightarrow[\mathsf{HTTPS}]{\mathsf{StoreResult}} \mathsf{MozaikObelisk}$ \label{state:fheprocessing:store} 
\EndFor
\end{algorithmic}
\end{algorithm}

\section{Use Case Description and Requirements}\label{sec:use-case}

The MOZAIK platform enables the collection and processing of various sensor and IoT data, without making any assumptions about the structure, format, or source of the data. To evaluate the capabilities of our proposed PoC system architecture and demonstrate its practical feasibility, we implement a realistic use case on top of MOZAIK.

\subsection{The Heartbeat Analysis Use Case}\label{sec:use-case:intro}
In our selected use case, users wear heartbeat sensors that continuously measure electrocardiogram (ECG) data. These sensors can be built into a wearable, for example, a smartwatch or other fitness trackers.
In this context, the service offered by the platform is continuous monitoring and diagnosis of common heartbeat anomalies that can be extracted from a single lead signal of an ECG.
Thus the user's device continuously sends the data points to the platform, which stores them and returns diagnosis predictions to the user. The predictions are computed using a machine learning model with pre-trained parameters. The model is provided by a third party, the so-called model provider.

During a 12-hour measurement period, approximately 21.5 MiB of heartbeat data is collected, while the model parameters for prediction require about 70 KiB. This storage demand may already be substantial for resource-constrained IoT devices. Moreover, such devices often have limited CPU processing capabilities. To address this, we outsource both heartbeat data storage and prediction computation to the cloud. Additionally, the model provider may be reluctant to deploy the prediction model on a user's device to protect their intellectual property and not to reveal business-related model parameters.

This ECG data is clearly sensitive and personal, requiring the need for a privacy-preserving storage and processing solution. Implementing this use case on top of the MOZAIK platform will keep the user's heartbeat data inaccessible from both the platform and model provider, while at the same time keeping the model parameters private from both the platform and the user. The only information leaked to the platform is the model architecture.

\subsection{Dataset and Model}\label{sec:use-case:dataset}
\begin{figure}
    \centering
    \begin{subfigure}[b]{0.49\linewidth}
        \centering
        \fontsize{8.5pt}{9.5pt}\selectfont
        \begin{tabular}{cl}
            \toprule
             Class & ECG annotation \\
             \midrule
             \multirow{4}{*}{N} & \textbullet~Normal \\
                                & \textbullet~Left/Right bundle branch block \\
                                & \textbullet~Atrial escape \\
                                & \textbullet~Nodal escape \\
             \midrule
             \multirow{4}{*}{S} & \textbullet~Atrial premature \\
                                & \textbullet~Aberrant atrial premature \\
                                & \textbullet~Nodal premature \\
                                & \textbullet~Supra-ventricular premature \\
             \midrule
             \multirow{2}{*}{V} & \textbullet~Premature ventricular contraction \\
                                & \textbullet~Ventricular escape \\
             \midrule
             F                  & \textbullet~Fusion of ventricular and normal \\
             \midrule
             \multirow{3}{*}{Q} & \textbullet~Paced \\
                                & \textbullet~Fusion of paced and normal \\
                                & \textbullet~Unclassifiable \\
             \bottomrule
        \end{tabular}
        \caption{Mapping of AAMI EC57 classes and heartbeat annotations (from~\cite{ECGDataset}).}
    \end{subfigure}
    \hfill
    \begin{subfigure}[b]{0.49\linewidth}
        \centering
        \includegraphics[width=\linewidth]{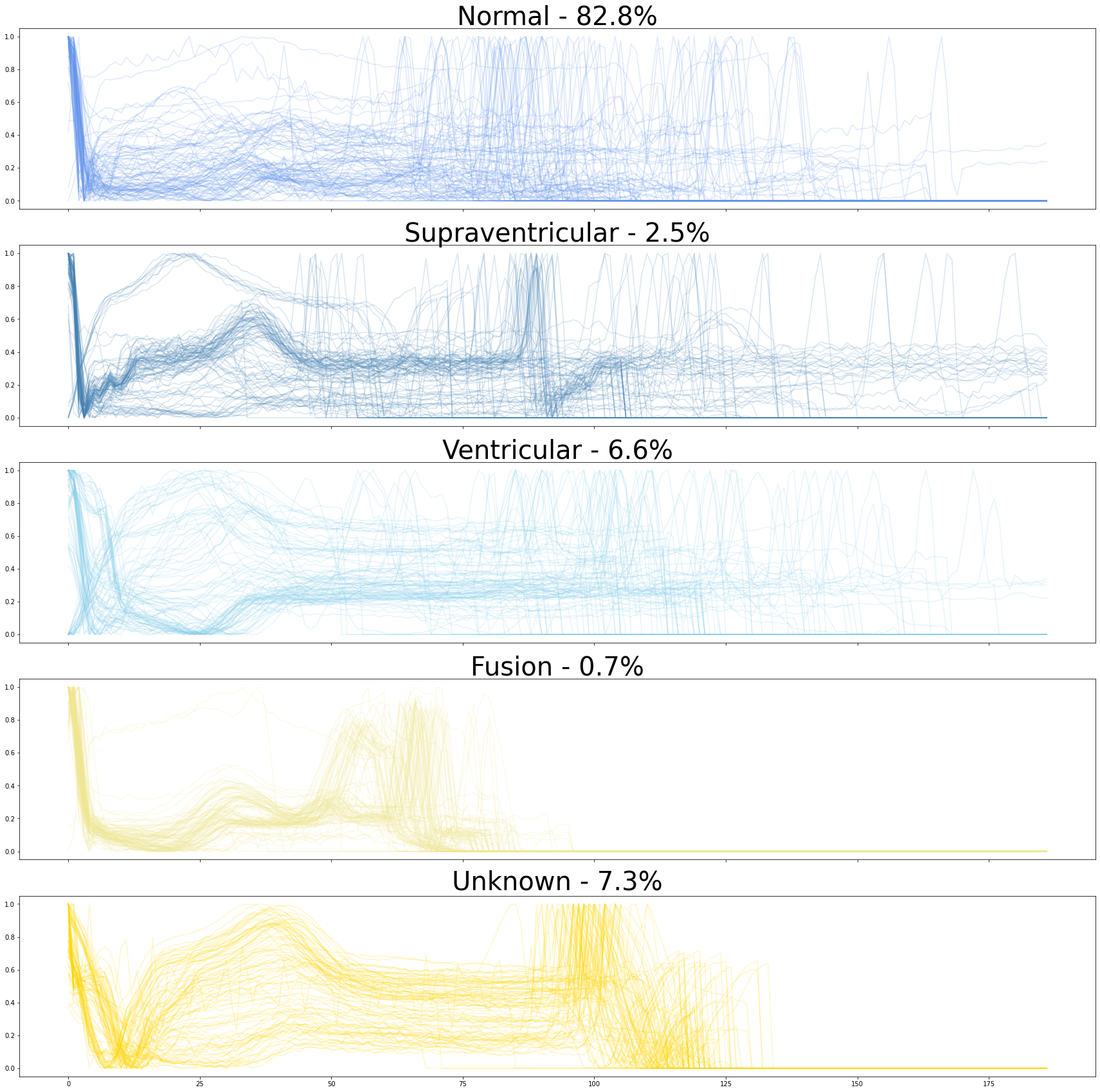}
        \caption{Visualization of the dataset samples for each class.}
    \end{subfigure}
    \caption{Summary of annotated classes of the heartbeat classification dataset.}
    \label{fig:use-case:heartbeat-dataset}
\end{figure}

In this work, we do not aim to improve data analysis techniques to enable diagnosis of common heartbeat anomalies but instead use an existing annotated dataset to showcase the processing part of the platform. We utilize the ECG heartbeat classification dataset by Kachuee et al.~\cite{ECGDataset}, which unifies the ECG lead II signal of the MIT-BIH Arrhytmia dataset~\cite{MIT-BIHDataset} with the PTB Diagnostic ECG database~\cite{PTBDataset}. The raw, single-lead ECG signal is preprocessed to extract all single heartbeats with surrounding signal of $1.2\times$ the length of the median heartbeat duration of a ten-second window. The signal is further normalized. Each heartbeat is classified into one of five classes according to the Association for the Advancement of Medical Instrumentation (AAMI) EC57 standard~\cite{AAMI-classes}. Figure~\ref{fig:use-case:heartbeat-dataset} shows (a) the five classes and (b) samples of each class. Each sample consists of 187 data points and represents a duration of approximately 1.5 seconds.

\begin{figure}
    \centering
    \includegraphics[width=0.7\linewidth]{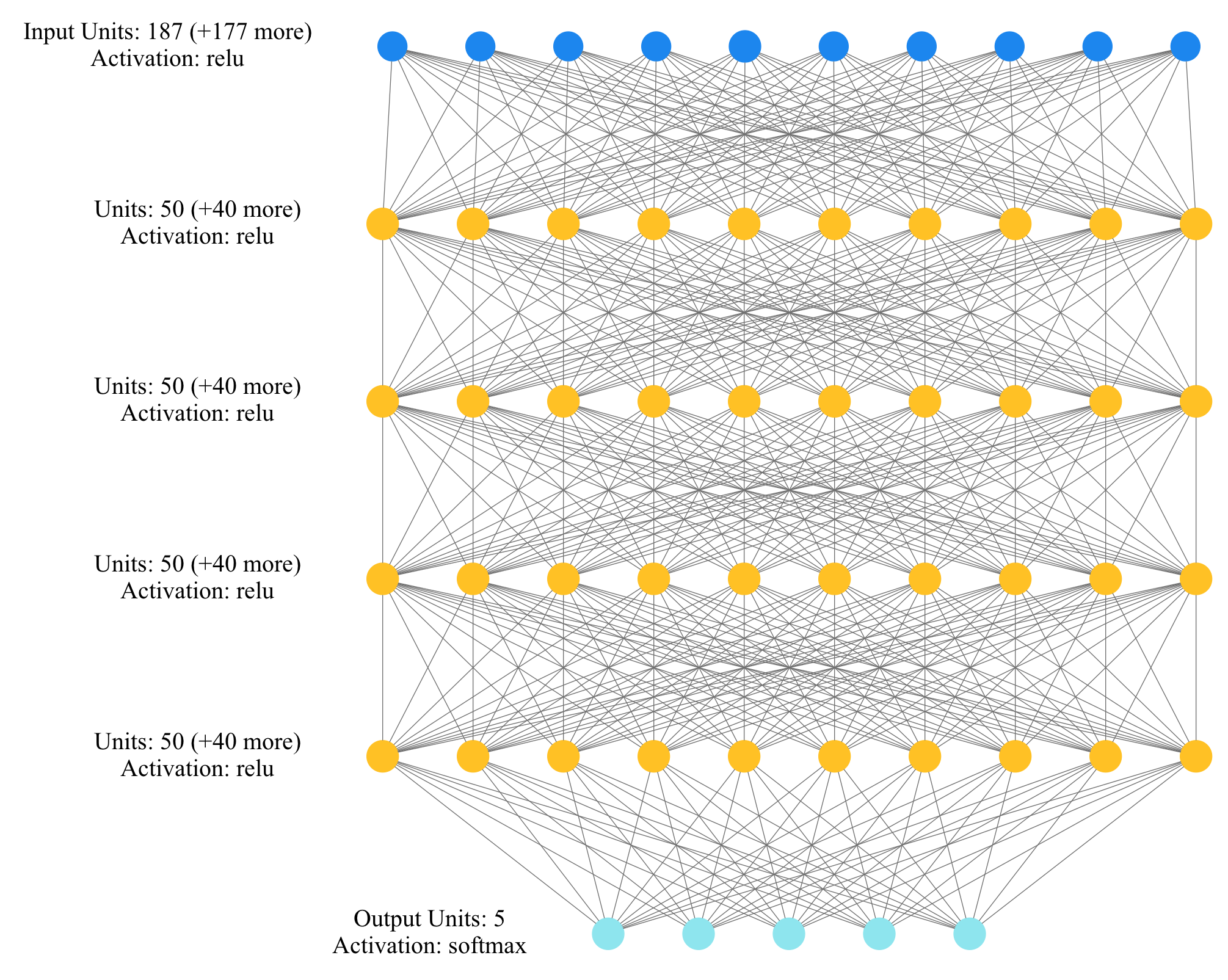}
    \caption{Visualization of the ECG heartbeat classification model.}\label{fig:use-case:heartbeat-model}
\end{figure}

In our use-case implementation, the model provider employs a conceptually simpler machine learning model than the model of Kachuee et al.~\cite{ECGDataset}. More specifically, our model consists of four dense layers with 50 intermediary nodes and a  dense output layer with five nodes, as shown in Figure~\ref{fig:use-case:heartbeat-model}. The accuracy of the plaintext ECG heartbeat classification is 96\%.

\subsection{Use Case Requirements}\label{sec:use-case:requirements}
We identify the following requirements, derived from the platform's design goals (see Section~\ref{sec:design-goals}), to guide our implementation of the use case.

\begin{description}
    \item[Privacy.] 
    The sensor data collected from a user remains private throughout all interactions with the platform and its entire lifecycle, including during processing. No sensitive information about the data can be learned by other users, the storage platform, or the model provider.
    
    \item[User Control.] 
    The sensor devices, such as smartwatches or fitness trackers, remain under the user's control.

    \item[User Trust.]
    The user remains in control even during data processing. The user has to explicitly authorize processing on a selected set of sensor data points by a specific set of processors. Processing of data outside the selected set or involving an unauthorized processor is not possible. Additionally, the user determines the duration of data storage and can retrieve their data at any time.
        
    \item[Correctness of the Diagnosis.] The diagnosis prediction delivered to the user is as accurate as if the user had direct access to the same model using their plaintext data.
\end{description}

From these fundamental requirements, we define the following reasonable performance parameters that the use-case implementation should adhere to.
\begin{description}
    \item[Data Ingestion Frequency.] The user's device sends one sample every 10 seconds.
    \item[Analysis Latency.] The platform makes an analysis result available to the user within 10 seconds.
\end{description}

As described in Section~\ref{sec:use-case:dataset}, preprocessing the raw ECG signal yields a variable number of `single' heartbeats depending on the user's heart rate. For example, at a resting heart rate of 80 beats per minute (bpm), approximately 13--14 beats are extracted from a ten-second window, whereas a rate of 150 bpm during exercise yields around 25 beats. However, since arrhythmias are typically characterized by patterns present in multiple atypical heartbeats, sending all beats within the ten-second window offers limited additional value compared to sending a fixed number. We therefore fix the number of beats in the ten-second window to one, allowing the sensor to arbitrarily select one representative beat from each window.

\section{Use Case Implementation Details}\label{sec:implementation}

The following section builds upon the general design and implementation details described in Section~\ref{sec:design} and provides use-case-specific implementation details for the use case described in Section~\ref{sec:use-case}. 

\subsection{Data Collection and Encryption at the Source}\label{sec:implementation:source}
Our prototype implementation emulates an IoT sensor device by reading a random sample from the ECG dataset described in Section~\ref{sec:use-case:dataset}. This sample is then encrypted using AES-GCM with AES-128 as the underlying block cipher, as outlined in Section~\ref{sec:design:details:encryption}, and then transmitted to MOZAIK-Obelisk via the REST API for storage in Obelisk.

\subsection{Key Share Creation and Result Decryption}
Creation and encryption of key shares and decryption of analysis results are performed in the user's browser through the web dashboard. To leverage standard implementations of cryptographic primitives provided by the browser, we chose to use RSA-OAEP as public-key encryption scheme and SHA-256 as the hash function, based on the limited options available via the Web Crypto API~\cite{noauthor_webcrypto_api_2024}.

\subsection{Data Processing}\label{sec:implementation:data-processing}
The following extends the general data processing implementation of Section~\ref{sec:design:details:coed} with the use-case-specific implementation details for both supported processing methods, MPC and FHE.

\subsubsection{Data Processing in MPC}\label{sec:implementation:data-processing:mpc}
Our data processing implementation in MPC comprises two main components, namely distributed decryption and encryption protocols, and a secure inference protocol. These MPC protocols operate on three servers in the honest majority setting, ensuring security against a malicious adversary with the possibility of abort. In the following, we detail each individual MPC component.

\paragraph{Distributed Decryption}
This step transforms the public ciphertext $\ctsymm_d^k$ into $\rssa{d}$ using shares of the encryption key $k_i$.
The MPC parties receive $N$, $\ctsymm_d^k$, $\mathsf{ID}_\mathsf{user}$, and obtain $k_i$ as described in Sections~\ref{sec:design:details:key-management} and~\ref{sec:design:details:coed}.
The parties compute $\Pi_\mathsf{DistDec}$ in Algorithm~\ref{alg:design:dist-dec} where $\ctsymm_d^k$, $N$ and $\mathsf{ID}_\mathsf{user}$ are public inputs, i.e., known to all parties, and the key share $k_i$ is the private input of party $P_i$.
We view $\Dec_{\rssb{k}}(\ctsymm_d^k, N, \mathsf{ID}_\mathsf{user} \| N)$ as a circuit where $\ctsymm_d^k$, $N$, and $\mathsf{ID}_\mathsf{user}$ are constants and $\rssb{k} = \rssb{k_1} + \rssb{k_2} + \rssb{k_3}$ is the encryption key that is reconstructed from the input $k_i$ by party $P_i$ but is still secret-shared.

If the concrete symmetric scheme is fixed, then more optimizations are possible to compute $\Dec$. In our use case, we use AES-GCM with AES-128 as block cipher.
We implement AES-128-GCM encryption and decryption in three-party replicated secret sharing by extending the MAESTRO framework~\cite{Maestro}. The framework implements oblivious AES-128 enciphering which we extend to the encrypt/decrypt operation of GCM~\cite{GCM}. 
This entails adding support for the GCM field $GF(2^{128})$ during the computation and verification phase. The batched multiplication verification in MAESTRO is implemented for $GF(2^{64})$. We show how to reduce the verification of a multiplication $u \cdot v = w$ with $u,v,w \in GF(2^{128})$ to two inner product checks of values in $GF(2^{64})$.
Let $P(X) = X^2 + \beta X + \gamma$ where $\beta, \gamma \in GF(2^{64})$ be an irreducible polynomial. We make use of the isomorphism $\phi: GF(2^{128}) \to (GF(2^{64}))^2$ to represent each element in $GF(2^{128})$ as an element of the degree-2 extension of $GF(2^{64})$ modulo $P(X)$. So, let $\phi(u) = aX + b$, $\phi(v) = cX + d$, and $\phi(w) = eX + f$, we write the multiplication as
\[
    (aX + b) \cdot (cX + d) \mod (X^2 + \beta X + \gamma) = (ad + bc + \beta ac)X + (bd + \gamma ac) \,.
\]
We can therefore verify the inner product checking two constraints: (1) $\rss{a} \cdot \rss{d} + \rss{b} \cdot \rss{c} + \beta \cdot \rss{a} \cdot \rss{c} = \rss{e}$ and (2) $\rss{b} \cdot \rss{d} + \gamma \cdot \rss{a} \cdot \rss{c} = \rss{f}$. We chose $\beta = 1, \gamma = 2^{61}$.

Note that the output shares $\rssb{d}$ of the distributed decryption are elements in $GF(2^8)$ (the AES field). Since the input to the MPC inference protocol is expected to be in arithmetic replicated secret sharing, the parties run the conversion protocol $\Pi_\mathsf{B2A}$ (Algorithm~\ref{alg:design:b2a}) to switch from boolean to arithmetic secret sharing.

\begin{algorithm}
\caption{Distributed decryption $\Pi_\mathsf{DistDec}$}\label{alg:design:dist-dec}
\begin{algorithmic}[1]
\Require Party index $i$, ciphertext $\ctsymm_d^k$, nonce $N$, user ID $\mathsf{ID}_\mathsf{user}$, key share $k_i$
\Ensure $\rssa{d}$
    \State $\rssb{k} := \rssb{k_1} + \rssb{k_2} + \rssb{k_3}$
    \State $\rssb{d} := \Dec_{\rssb{k}}(\ctsymm_d^k, N, \mathsf{ID}_\mathsf{user} \| N)$
    \State $\rssa{d} := \Pi_{\mathsf{B2A}}(\rssb{d})$ \hfill (see Algorithm~\ref{alg:design:b2a})
    \State \Return $\rssa{d}$
\end{algorithmic}
\end{algorithm}

\paragraph{Secure Inference}
Our MPC protocol for secure inference is based on arithmetic replicated secret sharing (see Definition~\ref{def:background:RSS}) and uses fixed-point precision with the parameters $\ell = 64$ and $f=8$. The precision of $8$ bits has been shown to be sufficient for maintaining the model's accuracy, achieving the same 96\% as in the plaintext case (see Section~\ref{sec:use-case:dataset}). Furthermore, by setting the precision to $8$ bits, we reduce the probability of large errors in a truncation protocol as detailed below. 

The protocol for secure inference is described in Algorithm~\ref{alg:design:infer}. It takes the ML model and the replicated secret shared input sample as input, and evaluates the forward pass of an $L$-layer deep neural network. Note that addition of secret shares can be done locally. On the contrary, to multiply two secret shares, the parties need to communicate to compute the cross-terms of the product, as described in Section~\ref{sec:background:mpc:rss}. 
We implement the ML inference on top of the MP-SPDZ framework~\cite{MPSPDZ}.
The protocol for generating Beaver triples in the offline phase is based on the preprocessing of Cramer et al.~\cite{cramer2018spd}. Activation layers involving non-linear operations are evaluated using binary circuits following the approach of ABY3~\cite{mohassel_aby3_2018}.

To accommodate the requirements outlined in Section~\ref{sec:use-case:requirements}, we use the MaSTer~\cite{zbudila_master_2025} truncation protocol to minimize the total inference time and maximize the throughput. While MaSTer is currently the most efficient truncation protocol for RSS, it exhibits probabilistic behavior, which may cause the honest parties to abort the computation unnecessarily. However, the probability of abort can be made small by setting the parameters appropriately; for example, choosing $\ell = 64$ and $f = 8$ would make this probability $<2^{-46}$, see MaSTer~\cite{zbudila_master_2025}. Note that MaSTer achieves its efficiency by operating under a new adversarial model. Recently, Zbudila et al.~\cite{Zbudila/cryptoeprint:2025/773} demonstrated that the MaSTer protocol is robust in realistic adversarial settings.

\begin{algorithm}
\caption{Secure inference $\Pi_\mathsf{Infer}$}
\label{alg:design:infer}
\begin{algorithmic}[1]
\Require Party index $i$, input $\rssa{d}$, model parameters $\rssa{W}_j, \rssa{b}_j$ $\forall j \in [L]$ where $L$ is the number of layers
\Ensure $\rssa{y}$
    \State $\rssa{y}_{\mathsf{input}} := \rssa{d}$ 
    \For{$j \in [L]$}
        \State $\rssa{y}_j := \Pi_{\mathsf{Actf}_j}(\rssa{y}_{j-1}\cdot\rssa{W}_j + \rssa{b})$
    \EndFor
    \State \Return $\rssa{y} := \rssa{y}_L$
\end{algorithmic}
\end{algorithm}

\paragraph{Distributed Encryption}
In the final step, the parties convert the prediction result $\rssa{y}$ into a ciphertext $\ctsymm_y^k$, $y$ encrypted under $k$.
The parties locally create the associated data $ad$ and nonce $N$ as
\begin{align*}
    ad & := \mathsf{ID}_\mathsf{user} \| \pk_1 \| \pk_2 \| \pk_3 \| \mathsf{ID}_\mathsf{analysis} \| \mathsf{type} \,, \\
    N & := \lfloor H(ad) \rfloor_{|N|} \,,
\end{align*}
where $H$ denotes a collision-resistant hash function and $\lfloor \cdot \rfloor_{|N|}$ is truncation to the nonce length, if necessary. The $\mathsf{ID}_\mathsf{analysis}$ is generated by the metadata store and $\mathsf{type}$ identifies the type of prediction.
By defining $ad$ in this way, we bind the prediction result to a particular request of the user and to the assigned MPC party pool. This excludes mix-and-match types of attacks by MOZAIK-Obelisk targeting the user.

Since the distributed encryption is based on the same techniques as distributed decryption described earlier, we first convert the prediction result $\rssa{y}$ from arithmetic into boolean sharing $\rssb{y}$ using the $\Pi_\mathsf{A2B}$ protocol (Algorithm~\ref{alg:design:a2b}) before running the distributed encryption. Protocol $\Pi_\mathsf{DistEnc}$ in Algorithm~\ref{alg:design:dist-enc} details this.

\begin{algorithm}
\caption{Distributed encryption $\Pi_\mathsf{DistEnc}$}
\label{alg:design:dist-enc}
\begin{algorithmic}[1]
\Require Party index $i$, inference output $\rssa{y}$, key share $k_i$, MPC party public keys $\pk_1, \pk_2, \pk_3$, analysis ID $\mathsf{ID}_\mathsf{analysis}$, user ID $\mathsf{ID}_\mathsf{user}$, prediction type $\mathsf{type}$
\Ensure $\ctsymm_y^k$
    \State $\rssb{k} := \rssb{k_1} + \rssb{k_2} + \rssb{k_3}$
    \State $ad :=\mathsf{ID}_\mathsf{user} \| \pk_1 \| \pk_2 \| \pk_3 \| \mathsf{ID}_\mathsf{analysis} \| \mathsf{type}$
    \State $N := \lfloor H(ad) \rfloor_{|N|}$
    \State $\rssb{y} := \Pi_{\mathsf{A2B}}(\rssa{y})$ \hfill (see Algorithm~\ref{alg:design:a2b})
    \State $\ctsymm_y^k := \Enc_{\rssb{k}}(\rssb{y}, N, ad)$
    \State \Return $\ctsymm_y^k$
\end{algorithmic}
\end{algorithm}

The MPC parties all obtain $\ctsymm_y^k$ after the distributed encryption protocol, which can be safely stored in Obelisk. Note that it might be tempting to have only one MPC party return the result to the database, however, this shortcut comes with a caveat.
Since MOZAIK-Obelisk, by design, does not know $k$, it cannot verify the validity of the ciphertext $\ctsymm_y^k$. Therefore, the single MPC party can arbitrarily modify the ciphertext causing decryption failures only at the moment the user accesses the results.
To address this denial of service attack vector, there are two options.
To \emph{detect} modification, a single MPC party sends $\ctsymm_y^k$ while the remaining only send a cryptographic checksum, e.g., a hash, of $\ctsymm_y^k$. MOZAIK-Obelisk can compare the received value with the received checksums and report the modification to the user and/or the remaining MPC parties.
Alternatively, to \emph{prevent} modification, all three MPC parties send $\ctsymm_y^k$. In that case, MOZAIK-Obelisk keeps the version of $\ctsymm_y^k$ where at least two parties agree.
We choose the second option to ensure a robust processing operation.

\subsubsection{Data Processing in FHE}\label{sec:implementation:data-processing:fhe}

In case of fully homomorphic encryption, we assume that ciphertexts are encrypted on the IoT device using the CKKS encryption scheme (see Section~\ref{sec:background:ckks}). All cryptographic operations are performed by using the OpenFHE~\cite{OPENFHE} library and, when requested, the decryption is performed in-browser using a WebAssembly port of OpenFHE. Furthermore, to interface with MOZAIK-Obelisk, we set up an HTTP server on the FHE engine that requests input ciphertexts and keys, or sends the inference results back.

In Algorithm~\ref{alg:fheinf} we describe the steps necessary to perform the FHE-based inference, leveraging the algorithms described in~\cite{HELIB} to compute the matrix-vector product. The inference algorithm makes use of two other subroutines, $\mathsf{ChebyshevApproximation}$ and $\mathsf{EvalPolynomial}$. Recall that SIMD-based FHE schemes are tuned towards efficient evaluation of polynomial operations over the encrypted vectors. However, the activation functions typically used in the evaluation of deep learning models are rarely polynomials. It is theoretically possible to compute the activation functions exactly, but in practice it will be more efficient to compute a polynomial approximation and evaluate the latter. The required polynomial is computed in the procedure $\mathsf{ChebyshevApproximation}$ and evaluated in $\mathsf{EvalPolynomial}$. Note that the latter simply relies on the $\FHEAdd$ and $\FHEMul$ functions, and is therefore not described in detail. Due to the use of approximations, the model's accuracy drops marginally from the plaintext baseline to 95\% (see Section \ref{sec:use-case:dataset}).

\begin{algorithm}
\caption{FHE-based inference}
\label{alg:fheinf}
\begin{algorithmic}[1]
\Require $\ct$: the ciphertext encoding the input vector of the inference.
\Require $\mathsf{KM}$: the auxiliary key material of the scheme.
\State Load $(W_i, b_i, F_i)_{i \in [0..k) \cap \mathbb{Z}}$, the weight matrices, biases and activation functions for all $k$ layers.
\State $\ct^0 := \ct$
\For{$i=0 \textbf{ to } i=k-1$}
    \State $\ct^i_{\mathsf{mm}} := \mathsf{FHEMatMul}(W_i, \ct^i)$ \hfill (see \cite{HELIB})
    \State $\ct^i_{\mathsf{lin}} := \FHEAdd(\ct^i_{\mathsf{mm}}, b_i)$
    \State $P_i(X) := \mathsf{ChebyshevApproximation}(F_i)$
    \State $\ct^{i+1} := \mathsf{EvalPolynomial}(\ct^i, P_i(X))$
\EndFor
\State \Return $\ct^k$
\end{algorithmic}
\end{algorithm}

\newpage
\section{Evaluation Results}\label{sec:evaluation}
This section evaluates the end-to-end privacy-preserving data storage and processing platform introduced in Section~\ref{sec:design}, focusing on the overhead it introduces compared to processing plaintext data without privacy protections. These evaluations are performed on the implemented use case described in Sections~\ref{sec:use-case} and~\ref{sec:implementation}. All evaluation results and scripts are available on GitHub~\cite{van_kenhove_github_nodate}.

\subsection{Experimental Setup}
Our experimental evaluation uses the following hardware.

\begin{description}
    \item[IoT Sensor and Gateway.] The IoT sensor device simulator that ingests samples into the system runs on a virtual machine server, and is supplied with 4 cores of a 2.40 GHz Intel(R) Xeon(R) E5645 CPU and 8 GB of RAM. The gateway runs on a separate virtual machine server with the same CPU type, but supplied with 8 cores and 16 GB of RAM.
    \item[MOZAIK-Obelisk.] All components encompassed within MOZAIK-Obelisk are deployed on a Kubernetes v1.29.1 cluster, consisting of one controller node and three worker nodes, all with 6 cores of a 2.40 GHz Intel(R) Xeon(R) Silver 4314 CPU and 32 GB of RAM.
    \item[MPC Parties.] We run the MPC protocol in a three-party setting on three virtual machine servers supplied with 4 cores of a 2.40 GHz Intel(R) Xeon(R) Silver 4314 CPU and 16 GB of RAM.
    \item[FHE Server.] We run the FHE inference on a virtual machine server supplied with 16 cores of a 2.10 GHz Intel(R) Xeon(R) Silver 4208 CPU and 64 GB of RAM.
\end{description}

The MOZAIK-Obelisk nodes and the MPC parties are interconnected via a 10 Gbit/s network link, achieving realistic data transfer rates of approximately 9.35 Gbit/s as measured with iPerf~\cite{noauthor_iperf_nodate}. The FHE server, on the other hand, is connected to MOZAIK-Obelisk using a 1 Gbit/s network link, achieving approximately 940 Mbit/s.

\paragraph{Experimental Setup for Micro-Benchmarks}
We provide micro-benchmark results for the data processing components using the same hardware setup described above. For comparison, we also include computation times for plaintext processing, that is, without applying privacy-preserving COED techniques. These plaintext processing experiments were conducted on a laptop equipped with a 24-core Intel(R) Core(TM) i7-12850HX CPU, 32 GB of RAM, and an RTX A2000 GPU with 8 GB of memory.

\subsection{Micro-Benchmarks of ML Inference}
In this section, we compare the inference runtimes of our deep neural network (DNN) in MPC and FHE to the corresponding plaintext runtimes when using the TensorFlow library~\cite{tensorflow2015-whitepaper,noauthor_tensorflow_nodate}. We present the results in Table~\ref{tab:micro-benchmarks:inference}.
The plaintext experiments are run in C++ on GPU using the cppflow~\cite{izquierdo2019cppflow} library utilizing the TensorFlow C API~\cite{noauthor_tensorflow_c_api_nodate}. The depicted times are averaged over ten experiments, measuring the total inference time, including setup and wrap-up procedures. Note that in the plaintext experiments, this entails data transfers from the CPU to the GPU and back, while in MPC, this involves the networking setup between the computing parties. We further compare the MPC protocol that is secure against a malicious (M) adversary to its semi-honest (SH) counterpart, highlighting the noticeable overhead introduced by the protocol that has stronger security guarantees. Finally, for the MPC experiments, we split the total running time into an offline (preprocessing) phase and an online (input-dependent) phase (see Section~\ref{sec:background:mpc}). Since the offline (preprocessing) phase does not depend on the input, the MPC parties could, in theory, perform it during idle times. This would speed up computation when a request arrives, as only the online phase would need to be executed.

The plaintext running times scale sublinearly with the batch size, with running times ranging from the order of microseconds to milliseconds. In contrast, the MPC overhead scales linearly with the number of required multiplications. For smaller batch sizes up to 16, the overhead introduced by MPC is minimal and has little practical impact compared to plaintext inference. However, as the batch size increases, the difference in performance becomes more pronounced. For a testing dataset of 10000 samples, running the inference using a maliciously secure MPC protocol can be up to five orders of magnitude slower than the plaintext version.

In the case of FHE-based inference, we implement single-sample and batched inference for our use case. In SIMD-based schemes, the maximal vector size that can be encrypted usually exceeds the input dimension of the model by several orders of magnitude. Hence, it is possible to perform the inference over batches by storing the concatenation of multiple vectors within the same ciphertext. Then, for a range of batch sizes, the runtime remains constant provided the amount of encoded vectors fits into the ciphertext.
However, unlike in MPC, it becomes a requirement that every vector in a batch is associated with the same user, as the keys must match.

\begin{table}
\centering
\caption{ML inference times for different batch sizes. The reported times compare the plaintext ML inference time with that of MPC in the semi-honest (SH) setting, MPC in the malicious (M) setting, and of FHE.}
\label{tab:micro-benchmarks:inference}
\begin{tabular}{llccc}
\toprule
\multicolumn{1}{c}{\multirow{2}{*}{Batch size}} & \multirow{2}{*}{} & \multicolumn{3}{c}{Time (s)} \\ 
\multicolumn{1}{c}{} &  & Preprocessing & Input-dependent & Total \\
\midrule
\multirow{4}{*}{1}  
 & Plaintext & - & - & 0.0006 \\
 & MPC (SH) & 0.004 & 0.078 & 0.082 \\
 & MPC (M) & 0.066 & 0.085 & 0.153 \\
 & FHE & - & - & 40 \\ \midrule
\multirow{4}{*}{16} 
 & Plaintext & - & - & 0.0009 \\
 & MPC (SH) & 0.015 & 0.127 & 0.145 \\
 & MPC (M) & 0.395 & 0.186 & 0.583 \\
 & FHE & - & - & 179.2 \\ \midrule
\multirow{4}{*}{32} 
 & Plaintext & - & - & 0.0009 \\
 & MPC (SH) & 0.031 & 0.171 & 0.204 \\
 & MPC (M) & 0.795 & 0.301 & 1.099 \\
 & FHE & - & - & 179.2 \\ \midrule
\multirow{4}{*}{64} 
 & Plaintext & - & - & 0.0009 \\
 & MPC (SH) & 0.049 & 0.292 & 0.343 \\
 & MPC (M) & 1.626 & 0.551 & 2.180 \\
 & FHE & - & - & 179.2 \\ \midrule
\multirow{4}{*}{128} 
 & Plaintext & - & - & 0.001 \\
 & MPC (SH) & 0.099 & 0.529 & 0.627 \\
 & MPC (M) & 3.183 & 0.965 & 4.151 \\
 & FHE & - & - & 179.2  \\ \midrule
 \multirow{4}{*}{256} 
 & Plaintext & - & - & 0.0015 \\
 & MPC (SH) & 0.200 & 0.990 & 1.194 \\
 & MPC (M) & 6.477 & 1.884 & 8.366 \\
 & FHE & - & - & 179.2 \\ \midrule
\multirow{4}{*}{512} 
 & Plaintext & - & - & 0.002 \\
 & MPC (SH) & 0.368 & 1.849 & 2.222 \\
 & MPC (M) & 13.086 & 3.901 & 16.993 \\
 & FHE & - & - & 358.7 \\ \midrule
\multirow{4}{*}{10000} 
 & Plaintext & - & - & 0.005 \\
 & MPC (SH) & 6.866 & 34.463 & 41.362 \\
 & MPC (M) & 249.587 & 71.196 & 320.858 \\
 & FHE & - & - & - \\ \bottomrule
\end{tabular}
\end{table}

\subsection{Communication Overhead}
In the following, the sizes of the relevant key material and ciphertexts required to achieve MOZAIK's end-to-end encryption design goal are detailed. These illustrate the communication overhead compared to plaintext data transfers.

For MPC, the symmetric key $k$ that is held by the user's IoT device is a standard AES-128-GCM key with a size of 16 bytes. Each encrypted sample consists of 1496 bytes and has a final size of 1524 bytes due to the added 12 bytes nonce and 16 bytes authentication tag. Each MPC party's RSA public key has a size of 256 bytes. Instead of secret sharing $k$, we secret share the key schedule of AES-128 for $k$, resulting in key shares of 176 bytes that need to be encrypted with RSA-OAEP. These three ciphertext key shares have a size of 256 bytes each, and each of these is intended for one of the three MPC parties. They are temporarily stored in MOZAIK-Obelisk's key store. The ciphertext of the analysis result, produced by performing the distributed encryption protocol, has a size of 56 bytes.

For FHE, there is a larger set of keys. Beyond the secret and public key, measuring 29 MiB and 71 MiB, respectively, the CKKS scheme requires additional keys to enable the homomorphic operations, totaling 26 GiB. We stress again that these additional keys can be publicly shared similarly to the public key. Finally, the input and output sample ciphertext have an identical size of 57 MiB.

\subsection{Performance of Ad Hoc Analysis}
To evaluate the performance of the proposed platform, we begin by assessing the computation latency of the ad hoc analysis processing mode by measuring the duration from the moment an analysis request is submitted by the user to the point at which the computation result is made available on the web dashboard. We define this as the end-to-end processing latency.

For each batch size, we construct an analysis request containing the corresponding number of data points for that batch size. Once the request is submitted, we wait until the computation results are stored in Obelisk, at which point they also become accessible to the user via the web dashboard. To calculate the end-to-end processing latency for each batch size, we use the timestamps of request submission and result storage that are recorded for each analysis in the metadata store. We repeat this process ten times for each batch size to compute the mean end-to-end processing latencies, which are detailed in Table~\ref{tab:evaluation:latency} for both MPC-based and FHE-based processing methods.

Note that FHE-based computation shows constant latency across batch sizes 1 through 8, and a higher constant latency for batch sizes 16 through 256. This is due to certain optimizations we include that only apply for smaller batch-sizes. Briefly, for smaller batch sizes, we can include multiple copies of the same sample in a single ciphertext. By exploiting this redundancy, we can decrease the required number of rotations and multiplications.

These results indicate that in ad hoc analysis mode, the MPC-based processing method in a malicious setting supports batch sizes of up to 240 while maintaining an end-to-end processing latency just below the ten-second requirement specified in Section~\ref{sec:use-case:requirements}. This highlights the cost of an end-to-end confidential data storage and processing solution, since the plaintext counterpart of a computation with batch size 240 would require less than 0.0015 seconds (see Table~\ref{tab:micro-benchmarks:inference}). The FHE-based processing method, on the other hand, does not meet the specified use-case requirements.

\begin{table}
    \centering
    \caption{Mean end-to-end (E2E) processing latency for varying batch sizes and the two evaluated COED methods in ad hoc analysis.}\label{tab:evaluation:latency}
    \begin{tabular}{crr}
        \toprule
        & \multicolumn{2}{c}{E2E Processing Latency (s)} \\
        Batch Size & MPC-based & FHE-based \\
        \midrule 
        1   & 1.059 & \multirow{4}{*}{41.017} \\
        2   & 1.075 & \\
        4   & 1.106 & \\
        8   & 1.354 & \\
        \cmidrule(l){3-3}
        16  & 1.931 & \multirow{16}{*}{180.217} \\
        32  & 2.784 & \\
        48  & 3.459 & \\
        64  & 4.020 & \\
        80  & 4.555 & \\
        96  & 4.962 & \\
        112 & 5.598 & \\
        128 & 6.245 & \\
        144 & 6.651 & \\
        160 & 7.149 & \\
        176 & 7.953 & \\
        192 & 8.479 & \\
        208 & 8.881 & \\
        224 & 9.280 & \\
        240 & 9.921 & \\
        256 & 10.507 & \\
        \bottomrule
    \end{tabular}
\end{table}

\subsection{Performance of Streaming}
\begin{figure}
    \centering
    \includegraphics[width=0.8\linewidth]{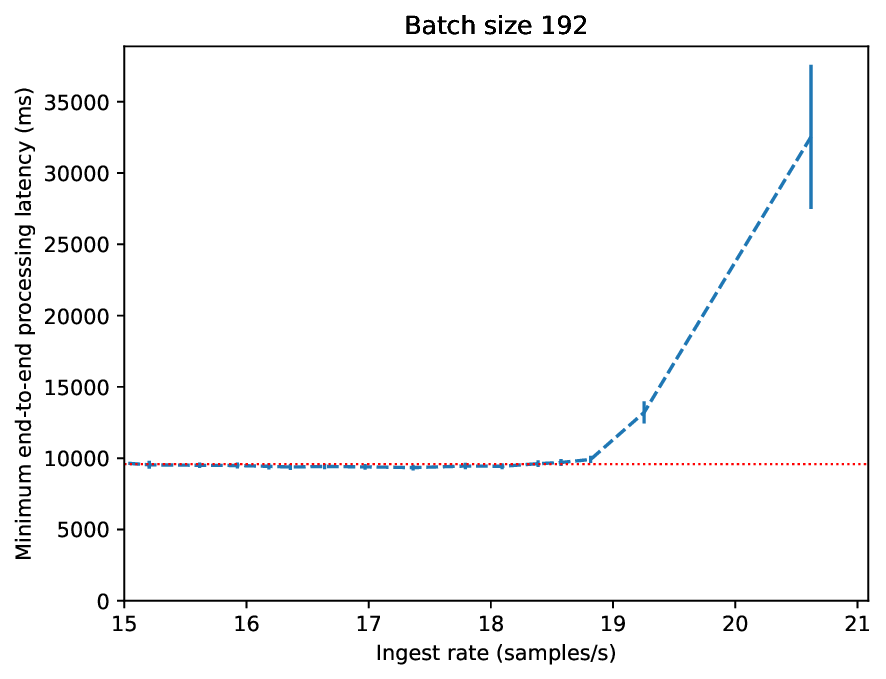}
    \caption{Minimum end-to-end processing latency for batch size 192 under a gradually increasing ingest rate. It is evident that the latency begins to rise significantly beyond an ingest rate of 18.8 samples/s, identifying this as the maximum sustainable ingest rate for this batch size. The red dotted horizontal line represents the mean latency of all processing runs up to and including this threshold.}\label{fig:evaluation:streamed_knee_graph}
\end{figure}

In streaming mode, on the other hand, data is continuously ingested at a certain ingest rate (in samples/s) and grouped into batches, with processing starting as soon as a batch is filled. Since ingesting data at a certain ingest rate requires time as well, we distinguish between two end-to-end processing latencies: the maximum end-to-end processing latency and the minimum end-to-end processing latency. The maximum end-to-end processing latency includes both the time required to fill a batch and the end-to-end processing latency itself, representing the longest possible delay an ingested data point may experience until it has been processed. The minimum end-to-end processing latency does not include the time required to fill a batch and reflects only the end-to-end processing latency. The data point that arrives first in a batch will experience the maximum end-to-end processing latency, whereas a data point that arrives last in a batch experiences the minimum end-to-end processing latency. Under normal load conditions, we expect that the minimum end-to-end processing latency for a batch of a certain size remains relatively constant. To evaluate the performance, we gradually increase the ingest rate for each batch size until we observe a significant rise in minimum end-to-end processing latency. The maximum ingest rate for a given batch size is the highest ingest rate at which the latency remains consistent with the latency observed at lower ingest rates. This method is visualized in Figure~\ref{fig:evaluation:streamed_knee_graph} for a batch size of 192, which has a minimal end-to-end processing latency below ten seconds under normal load, as an example. Here, the minimum end-to-end processing latency remains relatively stable up until an ingest rate of 18.8 samples/s. Ingesting more data beyond the maximum ingest rate would build up a queue with an ever-increasing length, eventually resulting in a failing system. Note that we only evaluate the streaming mode for the MPC-based processing method, since the FHE-based processing method, with an end-to-end processing latency of already 41.017 seconds for batch sizes one through eight, does not meet the ten-second analysis result requirement specified in Section~\ref{sec:use-case:requirements}.

Table~\ref{tab:evaluation:streaming} presents the evaluation results for MPC-based processing in streaming mode across varying batch sizes. Each row reports the maximum achieved ingest rate in samples/s for a certain batch size, the maximum and minimum end-to-end processing latencies at that ingest rate, and the number of supported data streams within a ten-second window at that ingest rate. The number of supported streams is derived from the ingest rate and reflects how many IoT sensor devices can be connected to the platform, assuming that each device ingests one data point every ten seconds, and that all connected devices ingest data points uniformly in the ten-second window. To meet the requirements outlined in Section~\ref{sec:use-case:requirements}, the maximum end-to-end processing latency must remain below ten seconds. Otherwise, some ingested data points will need to wait longer than ten seconds to retrieve their analysis result. The platform can therefore support an ingest rate of up to 13.841 samples/s at a batch size of 64, or 138 IoT sensor devices uniformly sending one data point each in a ten-second window. Here, the first data point in each batch will experience an average maximum wait time of 8.757 seconds, while the final data point in the batch will wait at least 4.205 seconds on average to retrieve their analysis result.

\begin{table}
    \centering
    \caption{Maximum ingest rate in samples/s achieved for streamed processing at certain batch sizes. The maximum and minimum end-to-end (E2E) processing latency at a given batch size respectively represents the longest and shortest time a data point might need to wait before its analysis result becomes available. The supported data streams reflect the maximum amount of IoT sensor devices that can be connected to the platform when performing streamed processing at a certain batch size.}\label{tab:evaluation:streaming}
    \begin{tabular}{crrrr}
        \toprule
        & \multicolumn{3}{c}{Streamed processing} &  \\
        Batch size & Samples/s & Max. E2E latency (s) & Min. E2E latency (s) & Data streams\\
        \midrule
        1   & 0.974  & 1.120 & 1.120  & 9   \\
        2   & 1.948  & 1.623 & 1.110  & 19  \\
        4   & 3.124  & 2.181 & 1.221  & 31  \\
        8   & 4.146  & 3.086 & 1.398  & 41  \\
        16  & 7.075  & 4.076 & 1.956  & 70  \\
        32  & 9.885  & 6.278 & 3.142  & 98  \\
        48  & 12.139 & 7.431 & 3.559  & 121 \\
        64  & 13.841 & 8.757 & 4.205  & 138 \\
        80  & 14.799 & 10.152 & 4.814  & 147 \\
        96  & 16.109 & 11.099 & 5.202  & 161 \\
        112 & 17.341 & 12.534 & 6.133  & 173 \\
        128 & 17.690 & 14.158 & 6.979  & 176 \\
        144 & 17.618 & 15.742 & 7.625  & 176 \\
        160 & 17.947 & 16.727 & 7.868  & 179 \\
        176 & 18.215 & 18.457 & 8.849  & 182 \\
        192 & 18.816 & 20.062 & 9.911  & 188 \\
        208 & 19.051 & 21.591 & 10.725 & 190 \\
        224 & 19.095 & 23.534 & 11.856 & 190 \\
        \bottomrule
    \end{tabular}
\end{table}

\section{Conclusion}\label{sec:conclusion}

As the volume of sensitive IoT data that require storage and analysis continues to grow, the need for an end-to-end secure and privacy-preserving storage and analytics system arose. This paper presents MOZAIK, an end-to-end confidential data storage and distributed processing solution for IoT-to-cloud scenarios. This work covers a detailed examination of the proposed architecture's design and implementation details. To validate MOZAIK's practical feasibility and better evaluate its capabilities, a realistic heartbeat anomaly detection use case was implemented. The platform ensures end-to-end data confidentiality by encrypting the data at the source before it leaves the user's control using a symmetric encryption key held by the user. The encrypted data is stored in Obelisk, a scalable, cloud-based time series database tailored for IoT data ingestion and querying. In order to provide secure data analytics and insights, two distinct COED techniques, MPC and FHE, are explored. These techniques allow privacy-preserving data processing, including ML inference, without revealing sensitive information about the data to the processors. The resulting encrypted analytics are stored back in Obelisk.

Our evaluation results demonstrate the feasibility of MOZAIK while also highlighting the substantial overhead associated with delivering end-to-end privacy-preserving data storage and analytics compared to conventional plaintext systems that offer no security or privacy. Despite the higher computational and performance costs, the strong privacy and security guarantees provided by MOZAIK make it a worthwhile trade-off in environments where data confidentiality is critical.

\paragraph{Trade-Offs Between MPC and FHE}
This work explores and implements both MPC and FHE COED techniques. Based on our experience, we briefly discuss the advantages and shortcomings of both technologies. On the one hand, MPC-based techniques offer sizable advantages in terms of performance. Even in the malicious setting, MPC-based inference outperforms FHE by a factor of 20 to 260 depending on the batch size. Furthermore, FHE-based techniques require additional key material on a per-user basis, which are of considerable size. The latter also impacts the computational efficiency: for any batch size, approximately 20 seconds are spent loading keys into RAM from disk. On the other hand, the deployment and management of the FHE pipeline is much simpler. Network traffic is only required at the very beginning and end of any computation, and no coordination between FHE servers is ever required. Due to the same independence, adding new FHE servers for horizontal scaling is straightforward, whereas scaling an MPC setup often involves complex coordination among parties, who must agree on legal frameworks and trust relationships. Furthermore, while the performance of MPC is bounded by the network properties, the performance of FHE is largely bound to the hardware and therefore can be improved by upgrading the latter. Considering the growing body of dedicated hardware acceleration for FHE~\cite{DPRIVE,NIOB}, it follows that the performance gap may be reduced or even closed in the future.

In general, our experiments suggest that employing FHE is preferable in scenarios where the network between MPC parties has low throughput or when the assumptions underlying the MPC security model are not believed to hold. However, when performance, particularly in terms of time or memory efficiency, is critical, MPC-based approaches will typically be more suitable. Moreover, MPC protocols offering malicious security are especially well suited for critical applications that require verifiable computation, an area where FHE currently falls short, unless combined with additional costly cryptographic mechanisms such as zero-knowledge proofs. Finally, a hybrid approach is also worth considering, where MPC parties simultaneously serve as FHE servers, allowing low-priority tasks to be delegated to the FHE components.

\backmatter






\section*{Declarations}

\bmhead{Funding} This work was supported by the Flemish Government through the \textit{Fonds Wetenschappelijk Onderzoek} (FWO), also known as the Research Foundation Flanders, under the SBO project MOZAIK with grant number S003321N. The project began on 1 May 2021 and concluded on 30 April 2025.

\bmhead{Competing Interests}
Filip De Turck, a listed author, is a member of the Editorial Advisory Board (EAB) of this Journal. The authors declare that they have no other competing interests.

\bmhead{Data, Materials, and Code Availability}
The complete MOZAIK architecture, design, implementation, and evaluation scripts and results are released as open-source software, bundled in the \url{https://github.com/MOZAIK-SBO/} GitHub organization.

\bmhead{Author Contributions}
\textit{Michiel Van Kenhove}, \textit{Erik Pohle}, \textit{Leonard Schild}, and \textit{Martin Zbudila} contributed equally to this work. Together, they designed, developed, and implemented the MOZAIK architecture over the past four years, and have all contributed substantially to the writing of this paper. \textit{Merlijn Sebrechts}, \textit{Filip De Turck}, \textit{Bruno Volckaert}, and \textit{Aysajan Abidin} supervised the project, providing valuable feedback and critical revisions to both the work and the manuscript. In addition, \textit{Aysajan Abidin} served as the project lead for MOZAIK, offering essential guidance that was instrumental to the project's successful completion. All authors reviewed and approved the final manuscript.


\bibliography{sn-bibliography}

\end{document}